\documentclass{optica-article}

\journal{opticajournal} 

\articletype{Research Article}

\usepackage{lineno}

\begin{document}

\title{Towards a mobile quantitative phase imaging microscope with smartphone phase-detection sensors}

\author{Xiangjiang Bao,\authormark{1,2,*} Zheng-da Hu,\authormark{3} Lucas Kreiss,\authormark{1} Josh Lerner,\authormark{1} and Roarke Horstmeyer\authormark{1,2}}

\address{\authormark{1}Department of Biomedical Engineering, Duke University, Durham, North Carolina 27708, USA\\
\authormark{2}Ramona Optics Inc., 1000 W Main St., Durham, North Carolina 27701, USA\\
\authormark{3}School of Science, Jiangnan University, Wuxi 214122, People’s Republic of China\\}

\email{\authormark{*}xiangjiang.bao@ramonaoptics.com}  


\begin{abstract*} 
Quantitative phase imaging (QPI) enables the visualization and quantitative extraction of optical phase information from transparent samples. However, conventional QPI techniques typically rely on multi-frame acquisition or complex interferometric optical setups. In this work, we propose \textit{Quad-Pixel Phase Gradient Imaging} (QP$^{2}$GI), a single-shot quantitative phase imaging method based on commercial quad-pixel phase detection autofocus (PDAF) sensors that are now commonly utilized in modern smartphones. PDAF sensors include an array of small microlenses, wherein each microlens covers a $2\times2$ pixel group on the sensor. The sample’s phase gradients induce focal spot displacements beneath each microlens, which in turn result in intensity imbalances across the four constituent pixels. By deriving the phase gradients of the sample from these imbalances, QP$^{2}$GI reconstructs quantitative phase maps from a single exposure. We establish a light-propagation model to describe this process and evaluate its performance in a customized microscopic system. Experiments demonstrate that quantitative phase maps of microbeads and biological specimens can be reconstructed from a single acquisition. Furthermore, low-coherence illumination improves robustness by suppressing coherence-related noise. These results reveal the potential of quad-pixel PDAF sensors as cost-effective platforms for single-frame QPI.

\end{abstract*}

\section{Introduction}
In traditional bright-field microscopy, the optical phase information of transparent samples cannot be directly measured. This is because standard image sensors only record light intensity and exhibit insensitivity to the phase of the optical field. Consequently, transparent or weakly scattering samples are challenging to observe, as they absorb negligible light and generate minimal intensity contrast. To address this limitation, various quantitative phase imaging (QPI) techniques~\cite{QPI1_nguyen_quantitative_2022,QPI2_Majeed} for measuring the optical phase of transparent objects.
Among these techniques, holography-based approaches\cite{HOLO_2_huang_quantitative_2024,HOLO1_mann_high-resolution_2005} recover the phase by interfering the sample beam with a coherent reference beam, enabling full-field quantitative reconstruction of the optical path difference. 
However, these interferometric methods typically require high temporal and spatial coherence of the illumination, precise reference-beam alignment, and vibration isolation, which limit their applicability in compact or cost-sensitive imaging systems. 

There are various non-interferometric methods for QPI that help overcome the above challenges. For example, QPI based on the transport of intensity equation (TIE)~\cite{6_sheppard_defocused_2004,7_bao_two_2019} can be implemented by introducing slight defocus during image acquisition; phase recovery is subsequently achieved by solving a partial differential equation that correlates intensity transport with phase gradients. Notably, this approach requires multi-frame acquisition and heavily relies on precise knowledge of both the defocus distance and the system’s optical transfer function. 

Other QPI methods are primarily based on structured or asymmetric illumination, such as quantitative differential phase contrast (DPC) imaging~\cite{1_lin_quantitative_2018,2_fan_optimal_2019,3_tian_quantitative_2015}.
These DPC implementations generally assume the weak-object approximation~\cite{5_hamilton_improved_1984} and derive both absorption and phase transfer functions under partially coherent illumination conditions via multi-frame acquisition. Fourier ptychographic microscopy (FPM)~\cite{FPM1_zheng_wide-field_2013,FPM2_ou_quantitative_2013} represents  another illumination-based QPI method which
synthesizes a large numerical aperture (NA) in Fourier space and reconstructs both amplitude and phase information by computationally fusing a sequence of low-resolution intensity images captured under varying illumination angles. 

To achieve single-snapshot QPI, a widely adopted strategy involves placing a microlens array (MLA) above the sensor pixels, as in a light field microscope (LFM)~\cite{9_levoy_light_2006,10_broxton_wave_2013,11_bimber_light-field_2019,12_mignard-debise_light-field_2015} or a Shack--Hartmann wavefront sensor (SHWS)~\cite{13_platt_history_2001,14_lane_wave-front_1992,15_primot_theoretical_2003}. In LFM, the MLA is positioned at the image plane to capture both the spatial and angular information of light rays emerging from the sample. The recorded light field enables numerical reconstruction of both amplitude and phase, at the expense of spatial resolution due to the division of the sensor into sub-apertures defined by individual microlenses. In contrast, SHWS utilizes the MLA to directly sample local wavefront slopes by measuring the displacement of focal spots formed by each microlens on the sensor. This local tilt information is then integrated to reconstruct the wavefront phase. SHWS is typically used in adaptive optics applications, for example, to measure ocular wavefronts in ophthalmology ~\cite{16_prieto_analysis_2000,17_wei_design_2010}, to inspect lens quality~\cite{18_jeong_measurement_2005,19_abdelazeem_characterization_2023}, or to detect wavefront aberrations caused by atmospheric turbulence~\cite{20_dayton_atmospheric_1992}. However, SHWS is less suitable for direct imaging of complex samples, as the sensor records a map of spot displacements rather than a two-dimensional image of the sample.

\begin{figure}[htbp]
\centering\includegraphics[width=12.5cm]{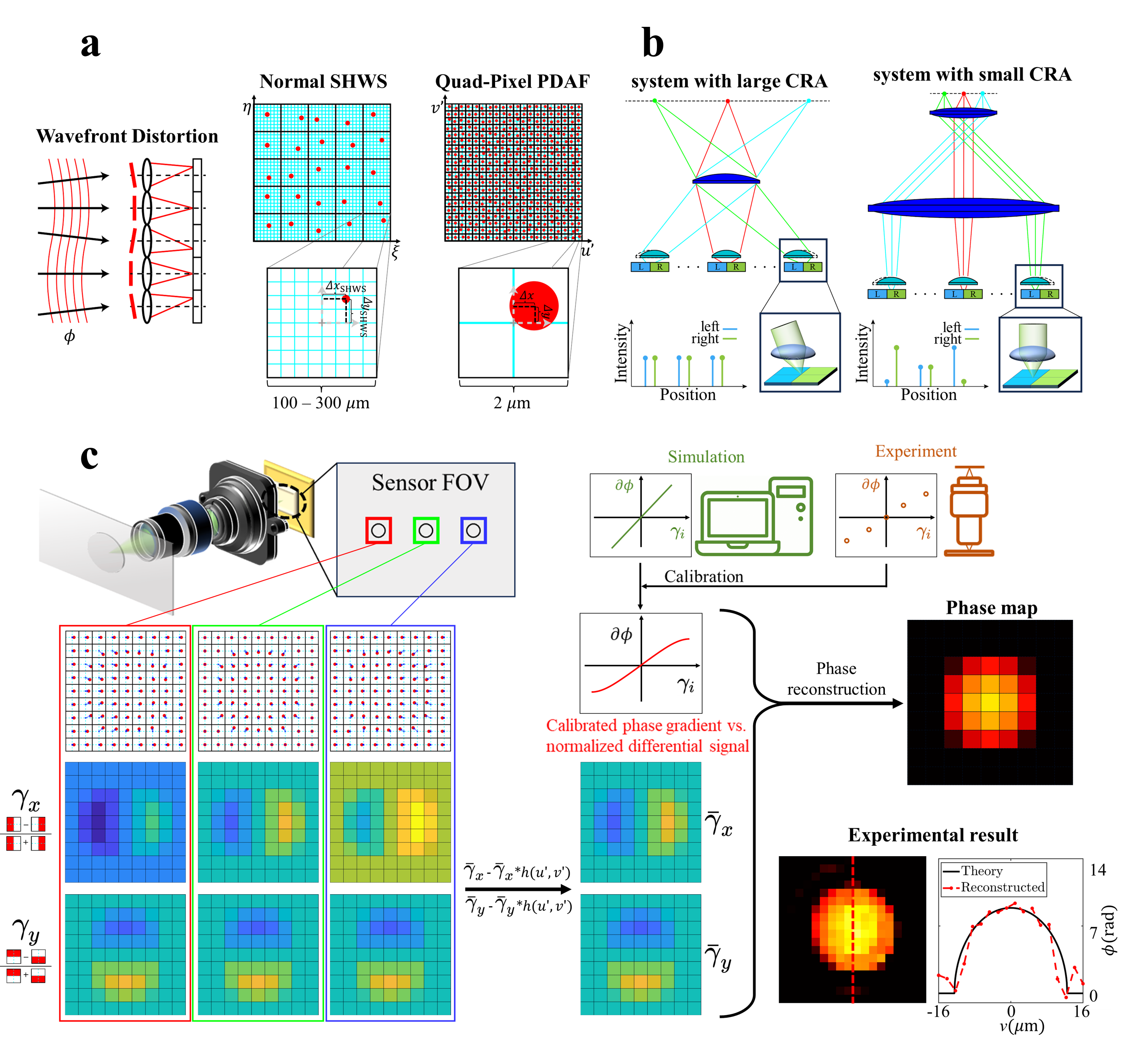}
\caption{Illustration of the phase detection principle and workflow of QP$^{2}$GI.(a) Comparison of the phase-detection principles between a conventional SHWS and a quad-pixel PDAF  sensor. 
(b) Illustration of the effect of chief-ray-angle (CRA) mismatch in an imaging system employing an MLA phase-detection sensor. The plots below show the responses of the left and right pixels at different sensor positions. 
(c) Workflow of QP$^{2}$GI and the reconstructed phase map of a $25~\mu\text{m}$-diameter microbead immersed in an environment with a refractive-index difference of $\Delta n = 0.03$.}
\label{fig1}
\end{figure}

Another imaging technique that employs an MLA is the phase-detection autofocus (PDAF) camera~\cite{21_kikuchi_125_2024,22_fukuda_compressed_nodate}. In 2021, OmniVision launched its first image sensor for smartphones (OV50A) with 100\% phase-detection coverage~\cite{23_noauthor_ov50a_nodate}. This sensor employs the quad-pixel PDAF technology, where each  $2\times2$ pixel group is covered by a single microlens, enabling all pixels to participate in phase detection while maintaining full imaging capability.
Prior research has demonstrated the potential of PDAF sensors for 3D imaging~\cite{24_jutamulia_phase_2022}. In this work, we explore  the potential of a quad-pixel PDAF sensor for QPI, and introduce \textit{Quad-Pixel Phase Gradient Imaging} (QP$^{2}$GI), a single-shot QPI method based on such sensors. Here, \textit{P}$^{2}$ simultaneously denotes the \textit{Pixel} and \textit{Phase} aspects of the method, as well as the $2\times2$ pixel structure employed for phase-gradient detection. In this configuration, the local optical phase gradient is inferred from estimating spot displacements under each microlens, which is similar to a conventional SHWS; however, unlike a typical SHWS that uses large microlenses (each mapped onto tens or hundreds of pixels), the quad-pixel PDAF sensor integrates a dense microlens array directly on the image sensor, with each microlens covering only four pixels (Fig.~\ref{fig1}a). This compact geometry effectively performs wavefront sampling at the image plane.  After pixel binning, the captured image retains high spatial resolution while encoding local optical phase-gradient information within each pixel group. The sensor features  a pixel width of $1.008$ $\mathrm{\mu}\text{m}$. After binning all pixels within each $2\times2$ pixel group, the pixel-limited spatial resolution is approximately $4~\mathrm{\mu}\text{m}/M$, where $M$ denotes the magnification of the imaging system. By leveraging this architecture, our work demonstrates that commercially available quad-pixel PDAF sensors--originally designed for smartphone cameras--have the potential to act as compact, low-cost, single-shot platforms for QPI.

\section{Theory and method}

\subsection{Phase Detection Principle of QP$^{2}$GI}
In this work, the quad-pixel PDAF sensor and the SHWS share a common fundamental principle for optical phase detection: both infer the local wavefront phase gradient by measuring the displacement of the focal spot formed beneath each microlens. In essence, a quad-pixel PDAF sensor can be regarded as a simplified SHWS, in which only four pixels are located under each microlens to detect the spot displacement (Fig.~\ref{fig1}a). Therefore, it is instructive to first revisit the basic principle of wavefront measurement in a conventional SHWS\cite{13_platt_history_2001}.

Ideally, in a common SHWS, if there is no wavefront aberration, the incident wavefront passes through each microlens and is focused onto the center of the corresponding sub-aperture. In this case, no spot shift is observed at the pixel plane. Here, and throughout this paper, the term \textit{pixel plane} refers specifically to the photosensitive surface. In contrast, when wavefront aberrations exist, the aberrated light is translated into measurable spot displacements (Fig.~\ref{fig1}a). We define $(\xi, \eta)$ as the coordinates on the pixel plane of a conventional SHWS, The lateral phase gradient  induced by aberrations can then be expressed as:
\begin{equation}
k_{\xi} = \frac{\partial \phi(\xi, \eta)}{\partial \xi}, \quad
k_{\eta} = \frac{\partial \phi(\xi, \eta)}{\partial \eta},
\label{eq1}
\end{equation}
where $k_{\xi}$ and $k_{\eta}$ are the transverse components of the wave vector $\mathbf{k}$, whose magnitude is $k = 2\pi / \bar{\lambda}$, with $\bar{\lambda}$ denoting the effective wavelength of light. Here, $\phi(\xi, \eta)$ is the optical phase of the incident wavefront in the microlens plane.  

Under the paraxial approximation, the wavefront tilt angles induced by the aberration for each sub-aperture can be calculated as
\begin{equation}
\theta_{\xi} \approx \frac{k_{\xi}}{k} = \frac{1}{k} \frac{\partial \phi(\xi, \eta)}{\partial \xi}, 
\quad
\theta_{\eta} \approx \frac{k_{\eta}}{k} = \frac{1}{k} \frac{\partial \phi(\xi, \eta)}{\partial \eta}.
\label{eq2}
\end{equation}

By applying the geometric relationship, the focal spot displacement is given by
\begin{equation}
\Delta x_{\mathrm{SHWS}} = f_l \theta_{\xi}, \quad
\Delta y_{\mathrm{SHWS}} = f_l \theta_{\eta},
\label{eq3}
\end{equation}
Here, the wavefront tilt angles $\theta_{\xi}$, $\theta_{\eta}$ represent the local slope of the wavefront within each sub-aperture, corresponding to the deviation of the propagation direction from the optical axis, and $f_l$ denotes the focal length of each microlens. Thus, the phase gradient information can be quantitatively obtained directly from the spot displacement, and the phase distribution of the light field can subsequently be reconstructed from these gradients in different directions.

From the previous derivation, it is essential to accurately determine the position of the focal spot center for a high-quality phase gradient measurement. For an SHWS with multiple pixels under each microlens, a simple way to determine the displacement of the focal spot by its center is to calculate the centroid of the intensity distribution in the focal plane~\cite{25_neal_shack-hartmann_2002,26_thomas_comparison_2006}:
\begin{equation}
\Delta x_{\mathrm{SHWS}} \approx X_c= \frac{\sum_i \sum_j X_i I(X_i, Y_j)}{\sum_i \sum_j I(X_i, Y_j)}, 
\quad
\Delta y_{\mathrm{SHWS}} \approx Y_c= \frac{\sum_i \sum_j Y_j I(X_i, Y_j)}{\sum_i \sum_j I(X_i, Y_j)},
\label{eq4}
\end{equation}
Here, $I(X_i, Y_j)$ denotes the intensity of the pixel located at position $(X_i, Y_j)$,
with $(i, j)$ representing the pixel indices within the sub-aperture beneath each microlens. 
The coordinate origin is defined at the geometric center of the pixel group beneath each microlens. 
$X_c$ and $Y_c$ denote the centroid coordinates of the focal spot (formed by each microlens) on 
the pixel plane. These coordinates represent the measured positions of the focal-spot centers, which correspond 
to the spot displacements $\Delta x_{\mathrm{SHWS}}$ and $\Delta y_{\mathrm{SHWS}}$ relative to the geometric center of the pixel group.

We now turn to the quad-pixel PDAF sensor. 
For optical phase detection with this sensor, it can be treated as a simplified SHWS where only  $2 \times 2$ pixels are located beneath each microlens. 
Each microlens samples a small local image region and focuses the collected light onto its corresponding $2 \times 2$ pixel group, where each pixel has a pitch of $a = 1.008~\mu\mathrm{m}$. 
When local phase gradients exist in the image, they induce angular tilts in the incident wavefront, which are then translated into lateral displacements of the focal spot relative to the pixel-group center.
We denote the measured intensities at the upper-left, upper-right, lower-left, and lower-right pixels as $I_{\mathrm{UL}}$, $I_{\mathrm{UR}}$, $I_{\mathrm{LL}}$, and $I_{\mathrm{LR}}$, respectively. 
Given that only $2 \times 2$ pixels are beneath each microlens, Eq.~(\ref{eq4}) reduces to the following expressions:
\begin{equation}
X_c = \frac{a}{2} \frac{(I_{\mathrm{LL}} + I_{\mathrm{UL}}) - (I_{\mathrm{LR}} + I_{\mathrm{UR}})}{I_{\mathrm{LL}} + I_{\mathrm{UL}} + I_{\mathrm{LR}} + I_{\mathrm{UR}}},
\quad
Y_c = \frac{a}{2} \frac{(I_{\mathrm{UR}} + I_{\mathrm{UL}}) - (I_{\mathrm{LR}} + I_{\mathrm{LL}})}{I_{\mathrm{UR}} + I_{\mathrm{UL}} + I_{\mathrm{LR}} + I_{\mathrm{LL}}}.
\label{eq5}
\end{equation}

As shown in Fig.~\ref{fig1}a, due to diffraction and microlens imperfections, the focal spot is not a perfect point, but instead it extends over a significant fraction of the $2 \times 2$ pixel region. Compared to conventional SHWS, the focal spot formed in a quad-pixel PDAF sensor occupies a relatively larger proportion of the area beneath each microlens. This spot size facilitates detectable intensity variations across the four pixels, thereby enabling phase-gradient sensing using only four pixels. 
However, this spot size also imposes limitations: Eq.~(\ref{eq5})  exhibits reduced accuracy, as 
$(X_c, Y_c)$ cannot accurately represent the actual displacement $\Delta x$ and $\Delta y$ of the extended focal spot, 
--owing to the inability of a $2 \times 2$ pixel group to fully sample the continuous light field projected onto this region.

We detail several strategies to address this limitation in the next subsections. Before proceeding, we first introduce two important variables for our analysis, $\gamma_x$ and $\gamma_y$, which denote the horizontal and vertical intensity differential signals, respectively. We define the directional intensities of the pixel group as \( I_{\mathrm{left}} = I_{\mathrm{UL}} + I_{\mathrm{LL}}, \;
I_{\mathrm{right}} = I_{\mathrm{UR}} + I_{\mathrm{LR}}, \;
I_{\mathrm{up}} = I_{\mathrm{UR}} + I_{\mathrm{UL}}, \;
I_{\mathrm{low}} = I_{\mathrm{LR}} + I_{\mathrm{LL}} \). With reference to Eq.~(\ref{eq5}), the differential signals are defined as follows:
\begin{equation}
\begin{gathered}
\gamma_x = \frac{I_{\mathrm{left}} - I_{\mathrm{right}}}{I_{\mathrm{left}} + I_{\mathrm{right}}}, \\
\gamma_y = \frac{I_{\mathrm{up}} - I_{\mathrm{low}}}{I_{\mathrm{up}} + I_{\mathrm{low}}}.
\end{gathered}
\label{eq6}
\end{equation}

\begin{figure}[htbp]
\centering\includegraphics[width=12.5cm]{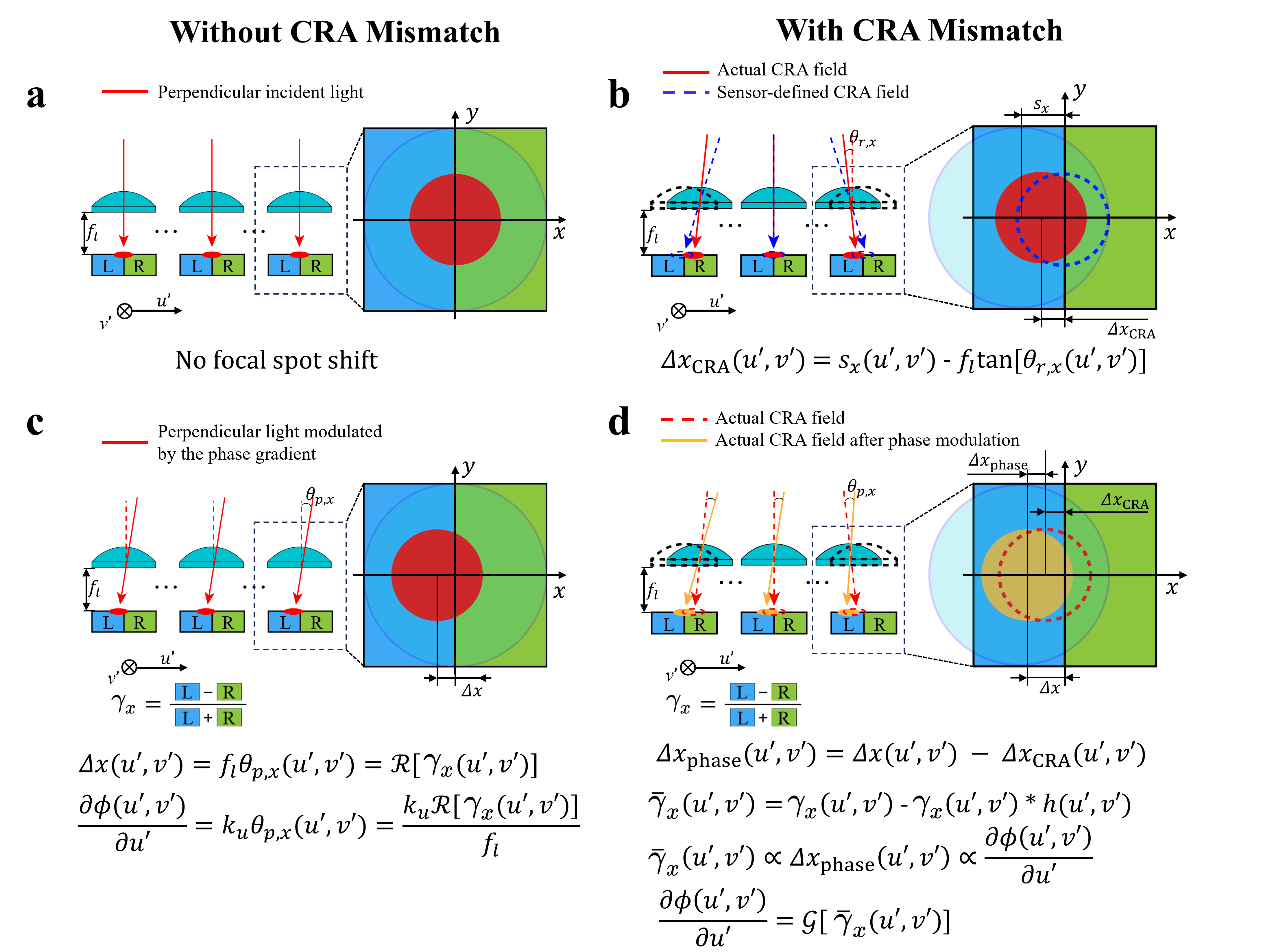}
\caption{Illustration of the correspondence between differential signals and object phase gradients.
(a) and (c) show the configuration without microlens shift, where the incident light is normal to the pixel plane in (a). In (c), a uniform phase gradient is applied to the incoming wavefront.
(b) and (d) show the configuration with an intentional microlens shift introduced to compensate for the CRA. However, the sensor-defined CRA does not match the actual CRA of the imaging system. In (d), the same phase gradient as in (c) is applied.}
\label{fig_illu}
\end{figure}

It is worth noting that the expressions for $\gamma_x$ and $\gamma_y$ in Eq.~(\ref{eq6}) closely resemble the standard formulation used to describe DPC signals under asymmetric illumination in Ref.~\cite{3_tian_quantitative_2015}, where $I_{\mathrm{left}}$, $I_{\mathrm{right}}$, $I_{\mathrm{up}}$, and $I_{\mathrm{low}}$  correspond to images acquired by blocking different halves of the illumination aperture.
Ideally, normalizing by the total intensity within each group of pixels makes the differential signals insensitive to uniform attenuation, ensuring that $\gamma_x$ and $\gamma_y$ respond primarily to phase-induced intensity asymmetries rather than amplitude variations related to absorption. Thus, these signals are directly correlated with the lateral displacement of the focal spot.
By taking a reference of Fig.~\ref{fig_illu}c, the phase gradient of the sample can be expressed as:
\begin{equation}
\begin{gathered}
\frac{\partial\phi(u',v')}{\partial u} = \frac{k_u\mathcal{R}[\gamma_x(u',v')]}{f_l},\\
\frac{\partial\phi(u',v')}{\partial v} = \frac{k_v\mathcal{R}[\gamma_y(u',v')]}{f_l},
\end{gathered}
\label{map}
\end{equation}
where $\mathcal{R}(\cdot)$ denotes a response function that maps the focal-spot displacements to the corresponding differential signals, and $(u', v')$ represent the global coordinates on the image plane. However, $\Delta x$ and $\Delta y$ are solely determined by the sample’s phase gradient provided that, in the absence of a sample, the focal spot formed beneath each microlens is perfectly centered on its corresponding pixel group (serving as the ideal reference position) as shown in Fig.~\ref{fig_illu}a. In what follows, we analyze the effect of  chief ray angle (CRA)  mismatch, which introduces an inherent focal-spot displacement even without a sample.

\subsection{Analysis of CRA mismatch}
CRA mismatch introduces a systematic spot displacement across the sensor, even in the absence of a sample. The quad-pixel PDAF sensor employed in this work was originally designed for smartphone photographic imaging. A key distinction between smartphone imaging and microscopic QPI resides in the CRA of the respective optical systems. While microscopic systems are typically engineered with a small CRA to maintain uniform focus and minimize aberrations, the PDAF sensor features a large CRA ($\approx 36^\circ$ at edge pixels, as specified by the manufacturer) to support a wide field of view (FOV). To accommodate this large CRA, the microlenses are intentionally offset with respect to the pixel groups.

As shown in Fig.~\ref{fig1}b, enables light incident at a steep angle to still focus near the center of the pixel group. In contrast, when the same sensor is employed in a microscopic system with a small CRA, off-axis incident light fails to focus accurately on the centers of the pixel groups. On the left side of the sensor, the focal spot formed by each microlens shifts rightward relative to that at the sensor center; whereas on the right side, the focal spot shifts leftward. The CRA mismatch therefore leads to nonuniform pixel responses within each group across the entire sensor FOV, manifesting as a background bias in the differential signals $\gamma_x$, as illustrated on the left side of Fig.~\ref{fig1}c. The background offset in $\gamma_x$ gradually increases from negative on the left side of the sensor to positive on the right side, reflecting the opposite directions of the focal-spot displacement.

Figure~\ref{fig_illu}b shows  the geometry of focal-spot displacement induced by CRA mismatch for each microlens along the horizontal direction. The red lines originate from the center of the imaging system’s exit pupil and thus represent the chief rays corresponding to each pixel group. The horizontal spot displacement of a pixel group induced by CRA mismatch is expressed as
\begin{equation}
\Delta x_\mathrm{CRA}(u',v') = s_x(u',v') - f_l \tan[\theta_{r,x}(u',v')],
\label{eq8}
\end{equation}
where $s_x$ denotes the horizontal microlens shift, and $\theta_{r,x}$ specifically refers to the horizontal CRA component for the corresponding pixel group. From Eq.~(\ref{eq8}), the horizontal spot displacement is determined by three parameters: the microlens shift $s_x$, the microlens focal length $f_l$, and the CRA $\theta_{r,x}$ associated with the pixel. Among these parameters, only $\theta_{r,x}$ is  determined by the incident light, while the other two are inherent to the sensor design. Additionally, $f_l$ affects the propagation of light after the microlens plane, thereby affecting the shape of the focal spot on the pixel plane.  microlenses deviate from the ideal thin-lens model, and the focal spot profile is sensitive to microlens imperfections,
which makes it unrealistic to fully simulate light propagation. Therefore, we investigated this relationship by experimentally determining $s_x$ and $f_l$, as detailed in the subsequent experimental section.

From the above analysis, the focal-spot displacement consists of two components: 
the CRA-induced displacement $\Delta x_{\mathrm{CRA}}$ and the displacement produced by the 
sample’s phase gradient $\Delta x_{\mathrm{phase}}$, as illustrated in Fig.~\ref{fig_illu}d. 
The resulting differential signals $\gamma_x$ and $\gamma_y$ therefore also contain contributions from 
both the CRA mismatch and the phase gradient of the sample. For samples containing fine structural features, the contribution from the CRA mismatch appears as a gradually varying low-frequency background across the sensor, while the contributions from the sample’s phase gradients manifest as high-frequency signals. To eliminate the background bias, background subtraction can be applied to the results in Eq.~(\ref{eq6}), which can be implemented 
through average filtering:
\begin{equation}
\begin{gathered}
\bar{\gamma}_x(u',v') = \gamma_x(u',v') - \gamma_x(u',v') * h(u',v'), \\
\bar{\gamma}_y(u',v') = \gamma_y(u',v') - \gamma_y(u',v') * h(u',v'). \\
\end{gathered}
\label{eq9}
\end{equation}
Here $h(u', v')$ represents the averaging filter kernel, whose size is adjusted according to the fine structural features of the sample. The symbol $*$ denotes the convolution operator.Alternatively, the background can be subtracted by referencing a sample-free image acquired under the same optical configuration. This process eliminates the low-frequency background and generates the normalized differential signals $\bar{\gamma}_x(u',v')$ and $\bar{\gamma}_y(u',v')$, 
which ideally originate solely from the sample’s phase gradient. Rather than the mapping in  Eq.~(\ref{map}), which correlates focal-spot displacement beneath each microlens with the differential signals, we can now establish a mapping that directly links the normalized differential signals to the sample’s phase gradients:
\begin{equation}
\begin{gathered}
\frac{\partial \phi(u',v')}{\partial u} = \mathcal{G}[\bar{\gamma}_x(u',v')],\\
\frac{\partial \phi(u',v')}{\partial v} = \mathcal{G}[\bar{\gamma}_y(u',v')],
\label{map_bar}
\end{gathered}
\end{equation}
where $\mathcal{G}(\cdot)$ denotes the system response function that maps focal-spot displacement to the sample’s actual phase gradients.

\subsection{Propagation model}
To establish the correspondence between the normalized differential signals $\bar{\gamma}_x$, $\bar{\gamma}_y$ and the actual optical phase gradients 
$\frac{\partial \phi(u,v)}{\partial u}$, $\frac{\partial \phi(u,v)}{\partial v}$, or equivalently to investigate the mapping $\mathcal{G}(\cdot)$ in Eq.~(\ref{map_bar}), a light-propagation model based on the Huygens–Fresnel principle is developed in this section.
In the experimental section, the simulation results generated from this model are further calibrated against experimental measurements, as illustrated on the right side of Fig.~\ref{fig1}. 

Prior to further analysis, we first define the coordinate system for the entire imaging configuration.
The global coordinates on the sample plane are denoted as $(u, v)$, and those on the image plane as $(u', v')$. 
For each $2\times2$ pixel group, the local coordinates within a pixel are represented by $(x, y)$, 
and the corresponding coordinates on the microlens as $(x', y')$. Consider a sample with thickness distribution $\beta(u,v)$ and refractive index $n_s$, immersed in an environment 
 with refractive index $n_0$. The optical phase distribution can be expressed as
\begin{equation}
\phi(u,v) = \frac{2\pi}{\lambda_0}(n_s-n_0)\beta(u,v) 
           = \frac{2\pi}{\lambda_0}\Delta n\,\beta(u,v),
\label{beta}
\end{equation}
where $\Delta n$ denotes the refractive index contrast between the sample and its surrounding environment.

Within this framework, the intensity distribution beneath each 
$2\times2$ pixel group is simulated using the Huygens–Fresnel integral 
\cite{27_goodman_statistical_2015}. The full mathematical derivation of the simulation is provided in Supplement~1, Section~S2. From the simulation, we obtain the directional integrated intensities 
$I^{\mathrm{sim}}_{\mathrm{left}}(u',v')$, 
$I^{\mathrm{sim}}_{\mathrm{right}}(u',v')$, 
$I^{\mathrm{sim}}_{\mathrm{up}}(u',v')$, and 
$I^{\mathrm{sim}}_{\mathrm{low}}(u',v')$, which serve as the simulated counterparts of the directional pixel intensities defined in 
Eq.~(\ref{eq6}).

In the simulation model, the microlens is treated as an ideal thin-lens phase element, 
and crosstalk effects arising from CRA mismatch are not explicitly included in the 
wave-propagation calculation. 
To account for these residual effects in a first-order manner, a proportional background 
term is incorporated by multiplying Eq.~(\ref{eq6}) by a constant term $B$, 
which accounts for the residual background signal:
\begin{equation}
\begin{aligned}
\gamma_x^{\mathrm{sim}}(u',v') = B \cdot \frac{I^{\mathrm{sim}}_{\mathrm{left}}(u',v') - I^{\mathrm{sim}}_{\mathrm{right}}(u',v')}{I^{\mathrm{sim}}_{\mathrm{left}}(u',v') + I^{\mathrm{sim}}_{\mathrm{right}}(u',v')}, 
\\
\gamma_y^{\mathrm{sim}}(u',v') = B \cdot \frac{I^{\mathrm{sim}}_{\mathrm{up}}(u',v') - I^{\mathrm{sim}}_{\mathrm{low}}(u',v')}{I^{\mathrm{sim}}_{\mathrm{up}}(u',v') + I^{\mathrm{sim}}_{\mathrm{low}}(u',v')}.
\end{aligned}
\label{eq12}
\end{equation}
While this remains a simplified assumption, it nonetheless enables the simulation to capture the sensor’s essential behavior and provide useful guidance for understanding its performance in the context of QPI.
A more accurate simulation can be performed using the finite-difference time-domain (FDTD) method~\cite{28_taflove_computational_2005}, 
by incorporating known features such as the microlens shape, microlens shift, 
and intrinsic stack layers of the pixels.

In the simulation, the wavefront incident on each pixel group is modeled as a tilted plane wave, 
with the overall tilt determined jointly by the sample-induced phase gradient and the incident CRA. A phase factor $\psi(x',y';u',v')$ is included in the model to represent the wavefront tilt 
introduced by the sample’s phase gradient. (See Eq. (S2))
By gradually changing this phase factor, we obtain $\mathcal{G}_{\mathrm{sim}}(\cdot)$, a simulated mapping 
that relates the normalized differential signals $\bar{\gamma}_x^{\mathrm{sim}}$ and 
$\bar{\gamma}_y^{\mathrm{sim}}$ to the corresponding phase gradients 
$\frac{\partial \phi(u,v)}{\partial u}$ and $\frac{\partial \phi(u,v)}{\partial v}$. Using a reference sample with known phase information, the 
experimentally calibrated mapping is then obtained through 
$\mathcal{G}(\cdot) \approx C_{\mathrm{cali}}\,\mathcal{G}_{\mathrm{sim}}(\cdot)$.
Now Eq.~(\ref{map_bar}) can be written as:
\begin{equation}
\begin{gathered}
\frac{\partial \phi(u,v)}{\partial u} \approx C_{\mathrm{cali}} \, \mathcal{G}_{\mathrm{sim}}[\bar{\gamma}_x(u',v')],\\
\frac{\partial \phi(u,v)}{\partial v} \approx C_{\mathrm{cali}} \, \mathcal{G}_{\mathrm{sim}}[\bar{\gamma}_y(u',v')],\\
\end{gathered}
\label{cali_eff}
\end{equation}
where $C_{\mathrm{cali}}$ denotes the calibration coefficient that scales the simulated mapping to match the experimentally measured phase gradients. In this work, microbeads with known size and refractive properties were used to determine the calibration coefficient.

Once the phase gradients are obtained from measured $\bar{\gamma}_x(u',v')$ and $\bar{\gamma}_y(u',v')$ by Eq.~(\ref{cali_eff}), the phase of the sample can be reconstructed by solving the Poisson equation 
in the frequency domain using Tikhonov regularization~\cite{29_noauthor_method_nodate}:
\begin{equation}
\phi(u,v) = 
\mathcal{F}^{-1}\!\left[
\frac{-j2\pi p\,{D}_u(p,q) - j2\pi q\,{D}_v(p,q)}
{(2\pi p)^2 + (2\pi q)^2 + \varepsilon}
\right],
\label{eq13}
\end{equation}
where $(p,q)$ denote the spatial-frequency coordinates, 
and ${D}_u(p,q)$ and ${D}_v(p,q)$ 
are the Fourier transforms of the measured phase gradients 
$\frac{\partial \phi(u,v)}{\partial u}$ and $\frac{\partial \phi(u,v)}{\partial v}$, respectively. 
The parameter $\varepsilon$ is a small regularization constant introduced to stabilize the inversion at low frequencies.

\subsection{Analysis of coherence}
To date,  we have established the principle and model of QP$^{2}$GI that illustrated in Fig.~\ref{fig1}c. In this section, we demonstrates that QP$^{2}$GI exhibits measurement repeatability and robustness when an illumination source with a coherence length exceeding the size of a single microlens is employed.
The analysis also confirms that within a specific range, a linear relationship exists between the normalized differential signal and focal-spot displacement, thereby ensuring the stability and consistency of phase-gradient measurements.

\begin{figure}[htbp]
\centering\includegraphics[width=12.5cm]{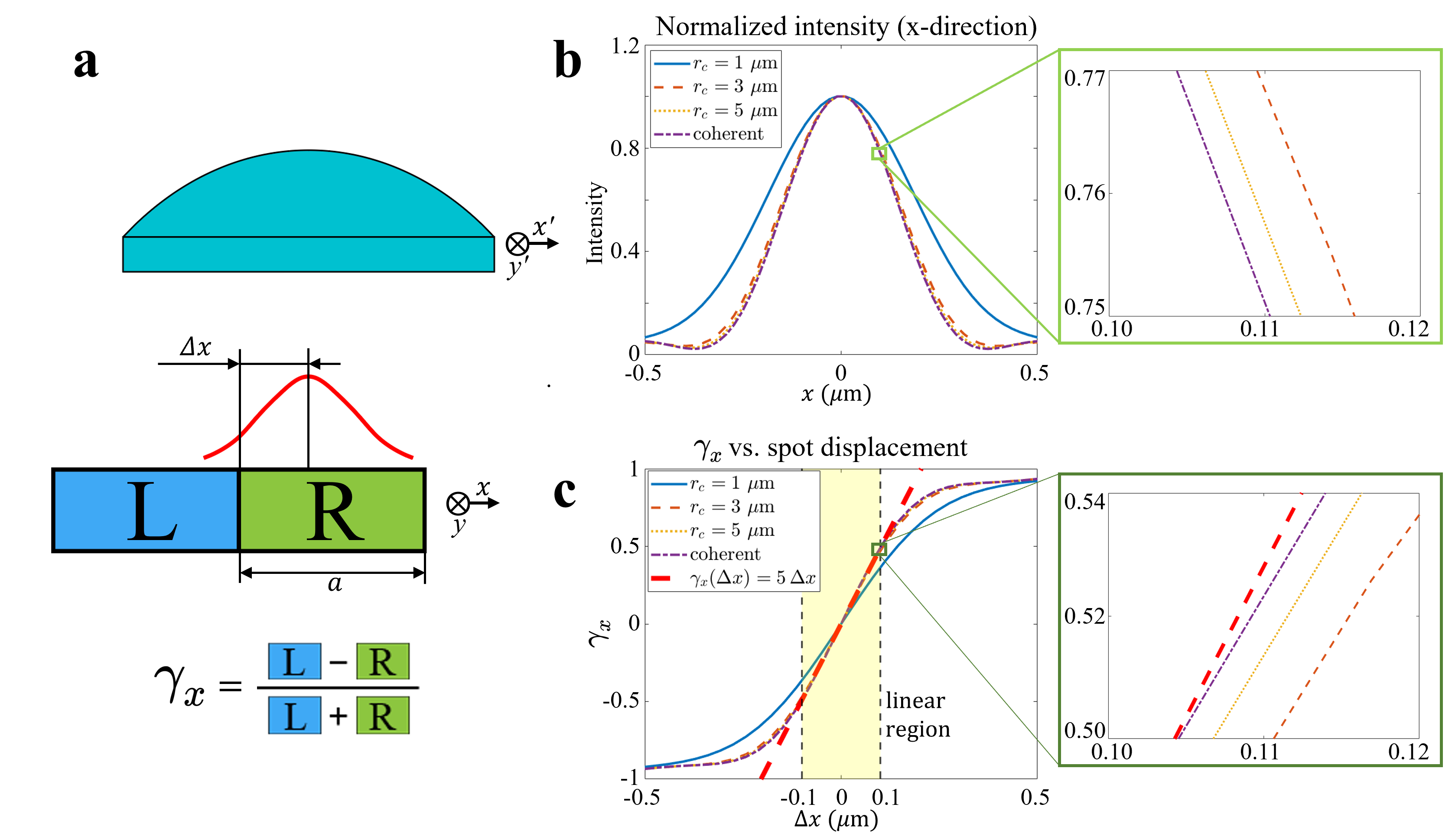}
\caption{Simulation results demonstrating the feasibility of optical phase measurement with a quad-pixel PDAF image sensor under different coherence lengths $r_c$. The incident light is perpendicular to the system and focused onto the center of a $2\times2$ pixel group. The microlens focal length is set to $f_l = 1~\mu\mathrm{m}$, and the microlens--pixel distance is assumed equal to $f_l$. (a)Schematic of the simulation layout. (b) Intensity distribution within one $2\times2$ pixel group along the $x$ direction. (c) Differential signal $\gamma_x$ as a function of the spot displacement $\Delta x$.}
\label{fig2}
\end{figure}

To study the effects introduced by illumination coherence, we simulated both the intensity 
distribution and the differential signals as functions of the focal-spot displacement beneath a 
microlens. For the intensity-distribution simulation, no microlens shift was applied and the 
incident light was assumed to be normal to the sensor surface. In contrast, when evaluating the 
differential signals as a function of the focal-spot displacement $\Delta x$, the focal spot was 
artificially shifted to emulate such displacement (Fig.~\ref{fig2}a). 
All simulations were carried out using the Huygens–Fresnel propagation model given in Section S2.

Figure~\ref{fig2}b illustrates the simulated beam profiles under varying illumination coherence conditions, with the distance between the microlens plane and pixel plane set to $1~\mu\text{m}$ in this simulation. This distance is estimated from the sensor’s CRA, based on the assumption that the lateral microlens shift for CRA compensation does not exceed half the size of the $2\times2$ pixel group, and The experimental procedure for obtaining the precise distance is detailed in the subsequent experimental section.
Strictly speaking, this distance ($\sim2\bar{\lambda}$, estimated from the center wavelength of the visible spectrum) lies at the boundary between the radiative and near-field regimes. However, since the microlens contains no sub-wavelength structures and the detector integrates the optical field over finite pixel areas, effectively acting as a spatial low-pass filter, the contribution of the remaining high-frequency near-field components is expected to be negligible in the actual experiment. 

Figure~\ref{fig2}b indicates that when the coherence length of the illumination exceeds the width of the pixel group ($\sim2~\mu\text{m}$), the light can be approximated as coherent during propagation. In practice, LED sources typically exhibit a relatively narrow bandwidth ($\sim30~\text{nm}$) compared to incandescent white light sources.
Thus, the coherence length can be estimated by $r_c = \lambda_0^2 / \Delta\lambda$, where $\lambda_0$ is the central wavelength  and $\Delta\lambda$ is the spectral bandwidth. 
Figure~\ref{fig2}c presents the relationship between the focal-spot displacement and the differential signal $\gamma_x$ for different coherence lengths. The curve of $\gamma_x$curve converges to the fully coherent case once the coherence length exceeds the pixel width. Moreover, in the regime of small spot displacements ($|\Delta x| < 0.1~\mu\text{m}$ in the simulation), the curve remains nearly linear. This provides a robust foundation for subsequent analysis, as the correspondence between focal-spot displacement and the differential signal remains stable under conditions of sufficient coherence. Conversely, this result also underscores a fundamental limitation of the phase-detection method: once the focal-spot displacement exceeds the linear regime, it becomes challenging to accurately quantify the relationship between focal-spot displacement and the differential signal.

\section{Experiment}

\subsection{Experimental setup}
\begin{figure}[htbp]
\centering\includegraphics[width=12.5cm]{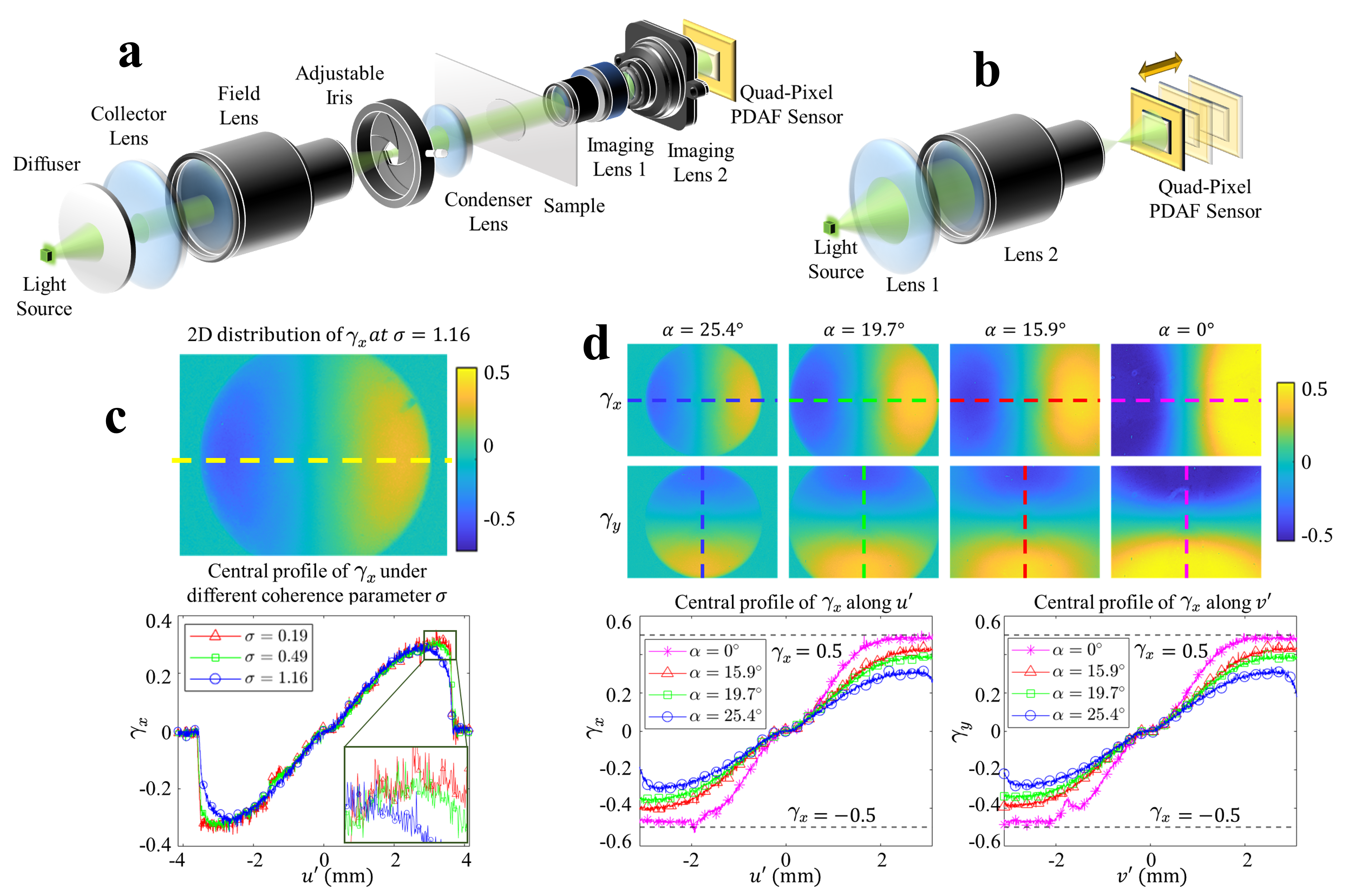}
\caption{Experimental setups. (a) 3D schematic diagram of the imaging setup. The light source is a $2~\mathrm{mm} \times 2~\mathrm{mm}$ LED centered at 530~nm with a bandwidth of $\sim30$~nm. Imaging lens~1 is a CCTV lens with a focal length of 2.8 mm and an f-number  of 2.2. Imaging lens~2 is a smartphone lens designed for a $1/1.55''$ sensor with a measured focal length of 8 mm and an $f$-number of 2. Together, these two lenses form a finite-conjugate imaging system with a measured CRA of~$26^\circ$. The system magnification $M$ is~2.3. The illumination coherence can be tuned by adjusting the iris. (b) 3D schematic diagram of the setup to simulate different CRAs. After collimation by Lens~1, the beam diameter is expanded to fully cover the clear aperture of Lens~2. The quad-pixel PDAF sensor is positioned beyond the focal plane of Lens~2 and illuminated by the diverging light. Lens~2 has an $f$-number of~1.4. (c) Differential signal $\gamma_x$ at different coherence parameters~$\sigma$, captured by the setup in~(a). (d) Horizontal differential signal $\gamma_x$ and vertical differential signal $\gamma_y$ captured at different incident angles~$\alpha$ with the setup in~(b). The maximum differential signal achieved by the system is approximately~0.5.}
\label{fig3}
\end{figure}

As elaborated earlier, the quad-pixel PDAF sensor’s large CRA ($\sim 36 ^\circ$) presents an inherent challenge for integration with a commercial microscopic system—since crosstalk induced by CRA mismatch precludes the reliable retrieval of the sample’s phase gradient. To evaluate the quad-pixel PDAF sensor’s QPI capability, we constructed a customized optical imaging system with a tailored illumination module (Fig.~\ref{fig3}a). In this system, an inverted CCTV lens (Imaging Lens 1) is positioned in front of a planar smartphone lens (Imaging Lens 2) to form a finite-conjugate microscope. Given that the CCTV lens has a larger f-number than the smartphone lens, it serves as the aperture-limiting element of the combined system.  Thus, the system’s effective NA  is constrained by the CCTV lens and can be approximated via its f-number as $0.23$. The CRA of this customized configuration was experimentally measured as 26° at the sensor edge. This value is determined by analyzing the correspondence between the illumination angle and the resulting differential signal (in the absence of a sample), as detailed in the section describing the experimental derivation of sensor parameters. Despite the remaining 10° CRA mismatch with the sensor, this system enables the reliable retrieval of the sample’s phase gradient.

The imaging-lens-limited maximum measurable phase gradient can be approximated as $
|\nabla \phi|_{\mathrm{max}} = \frac{2\pi}{\bar{\lambda}} \cdot \mathrm{NA}$,
which is calculated as 2.73~rad/$\mu$m at $\bar{\lambda}=530$~nm. However, since the system is constructed with two lenses in a finite-conjugate configuration, this value is expected to be smaller in practice, given that the sample is not positioned at the conjugate plane corresponding to the CCTV lens’s rated f-number.

An adjustable iris is  positioned at the conjugate plane of the diffuser, whereas the sample was placed at the conjugate plane corresponding to the diffuser’s Fourier plane. The illumination thus follows a Köhler-type configuration, in which the coherence of the illumination can be tuned by adjusting the iris aperture. In the following analysis, we describe the illumination coherence using the parameter $\sigma = \mathrm{NA}_{\mathrm{illum}} / \mathrm{NA}_{\mathrm{obj}}$, where $\mathrm{NA}_{\mathrm{illum}}$ and $\mathrm{NA}_{\mathrm{obj}}$ denote the NA of the illumination and the CCTV lens, respectively. This convention follows the definition used in QPI under partially coherent illumination, as adopted in DPC microscopy~\cite{3_tian_quantitative_2015}.

Figure~\ref{fig3}c demonstrates the image captured using the configuration described in Fig.~\ref{fig3}a  for different values of $\sigma$ in the absence of a sample. Due to the CRA mismatch, the differential signal $\gamma_x$ remains nonzero and increases with position even in the absence of a sample, representing the background signal caused by CRA mismatch, as shown in Fig.~\ref{fig1}c. It is observed that the $\gamma_x$ curves at different $\sigma$ largely overlap, except near the edge of the imaging FOV. This observation is consistent with the discussion of Fig.~\ref{fig2}, where the differential signal remains the same with varying coherence as long as the coherence length of the illumination exceeds the size of the pixel group, thereby confirming the repeatability of phase-gradient detection across different coherence levels. The observed decrease in  $\gamma_x$ at the FOV edge with decreasing $\sigma$ may be attributed to reduced spatial coherence in the edge region of the FOV.

\subsection{Experimental derivation of sensor parameters}

To study the effect of CRA mismatch, we built the setup shown in Fig.~\ref{fig3}b. A small light source was placed at the focal plane of a larger lens to collimate the beam, which was then refocused by another lens with an $f$-number of~1.4. Different CRA conditions of the imaging system can be simulated by translating the sensor along the optical axis. The focal spot formed by the second lens can thus be considered the center of the imaging system’s exit pupil. In the experiment, the illumination fully covers the clear aperture of the second lens; thus, the maximum divergence angle of the beam beyond the focal spot formed by the second lens is constrained by its $f$-number.

In Fig.~\ref{fig3}d, the differential signals $\gamma_x$ and $\gamma_y$ are measured using the configuration shown in Fig.~\ref{fig3}b for different incident angles~$\alpha$ in the absence of a sample. For the measurement of $\alpha = 0^\circ$, an additional lens is placed after Lens~2 to collimate the beam. The results show that $\gamma_x$ and $\gamma_y$, measured in the absence of a sample, decrease as the CRA mismatch is reduced. However, it is observed that  at $\alpha = 0^\circ$, the differential signals converge to~0.5, which is consistent with our assumption in Eq.~(\ref{eq12}). Owing to factors such as crosstalk induced by CRA mismatch and imperfections in microlens geometry, $\gamma_x$ and $\gamma_y$ cannot achieve the ideal value of 1.0 even when the focal spot is fully shifted onto the pixels on one side of a $2 \times 2$ pixel group.

To quantitatively estimate the effective focal length $f_l$ of the system, the experimentally measured differential signals $\gamma_x$ are compared with the theoretical predictions. Three experimental curves of differential signals across the sensor FOV, acquired under different incident angles $\alpha_1 = 15.9^\circ$, $\alpha_2 = 19.7^\circ$, and $\alpha_3 = 25.4^\circ$, are denoted as $\gamma_x^{(1)}(u')$, $\gamma_x^{(2)}(u')$, and $\gamma_x^{(3)}(u')$, respectively. The corresponding distances from the exit pupil to the sensor plane are labeled as $z_1$, $z_2$, and $z_3$.

Based on the theory presented above, we simulate $\gamma_x(\Delta x_\mathrm{CRA}; f_l)$ for different microlens focal lengths $f_l$ under normal illumination following a process similar to that used for Fig.~\ref{fig2}c, with the horizontal axis represented as the focal-spot displacement $\Delta x_\mathrm{CRA}$ in each $2 \times 2$ pixel group. The amplitudes of the $\gamma_x(\Delta x_\mathrm{CRA}; f_l)$ curves are scaled by a factor of~0.5. This scaling factor represents the background term $B$ in Eq.~(\ref{eq12}) and is determined from the experimental result in Fig.~\ref{fig3}b, in which $\gamma_x$ reaches a maximum of approximately~0.5. By comparing one of the experimental curves, $\gamma_x^{(2)}(u')$, with the corresponding theoretical curve $\gamma_x(\Delta x_\mathrm{CRA}; f_l)$, the reference focal spot displacement curves $\Delta x_\mathrm{CRA}^{(2)}(u'; f_l)$ for different focal lengths are determined. In practice, the curve $\gamma_x^{(2)}(u')$ is smoothed to derive the reference focal-spot displacement. From this displacement, the lateral shift of the microlens was calculated as:
\begin{equation}
s_x(u';f_l) = \Delta x_\mathrm{CRA}^{(i)}(u';f_l) + \frac{f_l}{z_i} u',
\label{eq14}
\end{equation}
where $z_i$ is the corresponding exit pupil distance from the sensor plane. This formula is derived from Eq.~(\ref{eq8}) by substituting $\tan(\theta_{r,x})$. The theoretical curves of the differential signals across the sensor FOV, $\hat{\gamma}_x^{(1)}(u'; f_l)$ and $\hat{\gamma}_x^{(3)}(u'; f_l)$, corresponding to the other two incident angles, are reconstructed from the derived $s_x(u'; f_l)$. Specifically, $\Delta x_\mathrm{CRA}^{(1)}(u'; f_l)$ and $\Delta x_\mathrm{CRA}^{(3)}(u'; f_l)$ are first calculated via Eq.~(\ref{eq14}), and then compared with the simulated $\gamma_x(\Delta x_\mathrm{CRA}; f_l)$ curve under the corresponding microlens focal lengths to obtain $\hat{\gamma}_x^{(1)}(u'; f_l)$ and $\hat{\gamma}_x^{(3)}(u'; f_l)$.

\begin{figure}[htbp]
\centering\includegraphics[width=10cm]{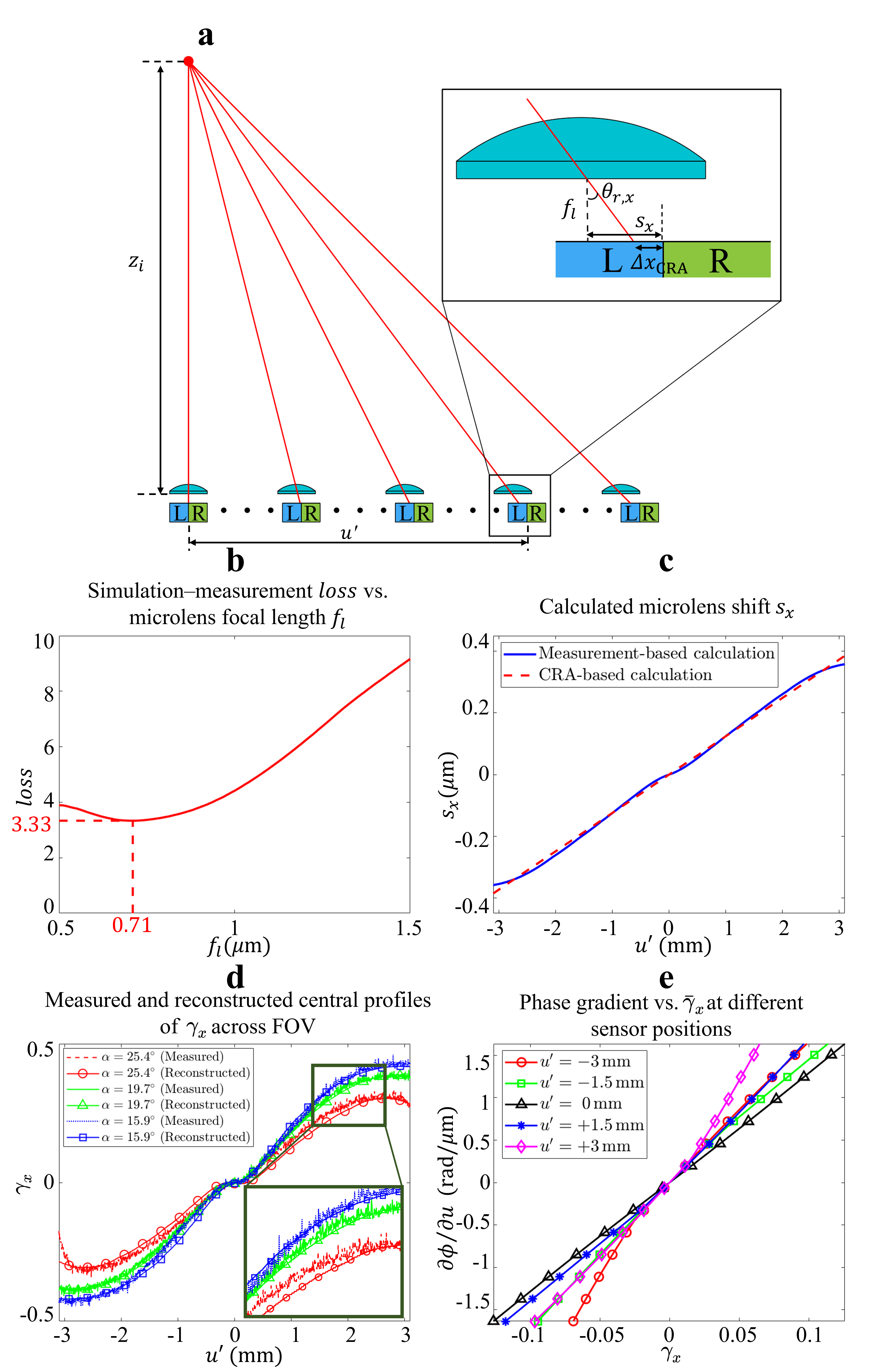}
\caption{Results of experimental data and simulations. (a) Geometric relationship between the microlens focal length $f_l$, local chief ray angle $\theta_{r,x}$, focal spot displacement $\Delta x_\mathrm{CRA}$, and microlens shift $s_x$. The red ray represents the ray passing through the center of the exit pupil of the imaging system. Due to CRA mismatch, the ray does not fall at the center of the pixel group. (b) Loss as a function of the microlens focal length, with $f_l = 0.71~\mu\mathrm{m}$ corresponding to the minimum loss. (c) Comparison of the microlens shift $s_x$ calculated from the sensor CRA with that the sensor CRA and that derived from the measured data. (d) Comparison of the differential signal $\gamma_x$ between measured data and reconstructed curves at different incident angles~$\alpha$ by applying the calculated microlens shift~$s_x$. (e) Relationship between the phase gradient $\partial\phi/\partial u$ and normalized differential signal $\bar{\gamma}_x$ at different sensor positions, obtained from simulations under the conditions $f_l = 0.71~\mu\mathrm{m}$ and $\alpha = 26^\circ$, scaled by a calibration factor of~0.5}
\label{fig4}
\end{figure}

Finally, the best-fit focal length $f_l$ is determined by minimizing the difference between the measured and simulated differential signals, expressed as the loss:
\begin{equation}
\mathrm{loss}(f_l) = \sum_{u'} \left\| \Gamma_x(u') - \hat{\Gamma}_x(u'; f_l) \right\|_2^2,
\label{eq15}
\end{equation}
where
\begin{equation}
\Gamma_x(u') =
\begin{bmatrix}
\gamma_x^{(1)}(u') \\
\gamma_x^{(2)}(u') \\
\gamma_x^{(3)}(u')
\end{bmatrix},
\quad
\hat{\Gamma}_x(u'; f_l) =
\begin{bmatrix}
\hat{\gamma}_x^{(1)}(u'; f_l) \\
\hat{\gamma}_x^{(2)}(u'; f_l) \\
\hat{\gamma}_x^{(3)}(u'; f_l)
\end{bmatrix}.
\end{equation}
The loss  attains its minimum at $f_l = 0.71~\mu\mathrm{m}$ (Fig.~\ref{fig4}b). this distance is relatively small, it is geometrically feasible for a sensor designed with a $36^\circ$ CRA and a pixel pitch of$1.008~\mu\mathrm{m}$. As $f_l$  increases, the required microlens shift increases—such that the microlens may no longer adequately cover the target pixel within its $2 \times 2$ pixel group. Back-side illumination (BSI) techniques \cite{30_holland_development_1997,31_taverni_front_2018} can reduce such distances by relocating the metal wiring layers to the back of the photosensitive region—thereby positioning the photosensitive region closer to the microlens.

The microlens shift $s_x(u';f_l)$ calculated from Eq.~(\ref{eq14}) as the blue solid line in Fig.~\ref{fig4}c. For comparison, the red dashed line denotes the microlens shift curve calculated directly from the sensor’s edge CRA, given by:
\begin{equation}
s_x(u';f_l) = \frac{f_l}{z_{\mathrm{CRA}}} u',
\label{eq16}
\end{equation}
where $z_{\mathrm{CRA}}$ represents the distance from the exit pupil to the sensor plane such that the edge light falls onto the center of the edge pixel group. As observed in Figs.~\ref{fig3}c, d, and Fig.~\ref{fig4}d, $\gamma_x$ appears flattened in the central region ($|u'| \lesssim 0.12~\mathrm{mm}$). This effect is likely attributed to two factors: first, the microlenses in the sensor’s central region are not shifted to match the behavior of a practical imaging system; second, the focal spot in Fig.~\ref{fig3}b is not a perfect point but has a finite geometric size, resulting in uniform light distribution in the central region. The blue solid curve in Fig.~\ref{fig4}c illustrates this feature, where $s_x$ exhibits a smaller slope in the central region. The reconstructed $\gamma_x(u')$ curves based on $f_l = 0.71~\mu\mathrm{m}$ are shown in Fig.~\ref{fig4}d, and the results exhibit relatively good consistency.

From the above discussion, the differential signals remain non-zero and increase across the sensor, even in the absence of a sample. This is regarded as a background offset induced by CRA mismatch between the optical system and the sensor that varies across the sensor FOV. The high-frequency signal components originating from the phase gradient can be isolated by subtracting this low-frequency background. this background is estimated by applying an average filter to the differential signals, as illustrated in Fig.~\ref{fig1}c. The resulting signals, which reflect only the phase gradient, are denoted as the normalized differential signals $\bar{\gamma}_x$ and $\bar{\gamma}_y$. Using the estimated $f_l$ and microlens shift $s_x$, simulations based on Eq.~(\ref{eq12}) are performed to establish the relationship between the actual phase gradient of the sample and the normalized differential signal at different sensor positions. As shown in Fig.~\ref{fig4}e,  The relationship between $\bar{\gamma}_x$ and the actual phase gradient of the sample is highly linear at the center of the sensor. Additionally, the same normalized differential signal corresponds to a higher phase gradient at positions farther from the center, indicating that the sensitivity of the sensor to phase gradients decreases with increasing distance from the center.

Although the model becomes less accurate toward the edges of the sensor since the simulated $\gamma_x(\Delta x_\mathrm{CRA}; f_l)$ curves are derived under perpendicular incidence, we can still regard it as providing important guidance for revealing the fundamental phase detection behavior of the sensor. Additionally, as the overall performance of this model is largely dependent on the accuracy of the estimated $\gamma_x(\Delta x_\mathrm{CRA}; f_l)$ curves,  a more precise simulation would require detailed information regarding the microlens geometry and arrangement.

\subsection{Phase detection and reconstruction results}

\begin{figure}[htbp]
\centering\includegraphics[width=12.5cm]{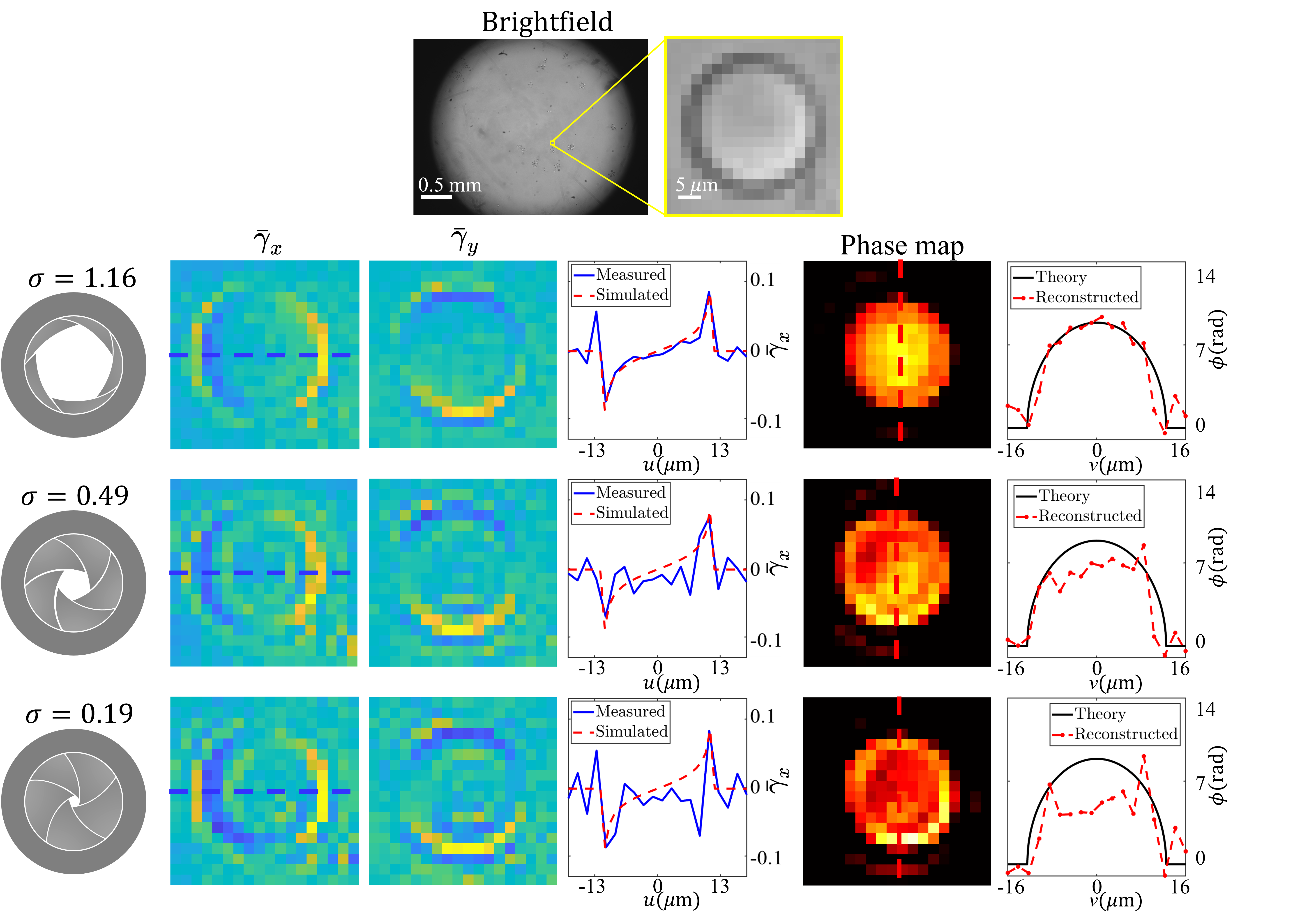}
\caption{Experimental results of microbeads captured at the center FOV of the sensor under different illumination coherence parameters $\sigma$. 
The grayscale image at the top shows a representative bright-field image obtained by binning all pixels within each $2\times2$ pixel group under the same microlens, along with the corresponding color channels. 
All microbeads are composed of the same material, with an average diameter of $25~\mu\text{m}$ and a refractive index of $n_\mathrm{s}=1.59$. 
The illumination coherence parameters for the three rows below are 1.16, 0.49, and 0.19, respectively. The simulated normalized differential signals (blue lines) are scaled by a calibration factor of 0.5 relative to the original simulation results.}
\label{fig5}
\end{figure}

To test and validate this approach, we image polystyrene microbeads (nominal diameter = 25~$\mu\mathrm{m}$, refractive index $n_s = 1.59$) immersed in an environment with $n_0 = 1.56$ using the imaging system shown in Fig.~\ref{fig3}a. Figure~\ref{fig5} shows the normalized differential signals and the corresponding reconstructed phase maps obtained under different illumination coherence parameters~$\sigma$. In the figure, the microbeads shown in the three rows correspond to the same microbead in the top bright-field image, captured by varying only the iris aperture. 

In this work, the simulated differential signals are derived from the propagation model based on Eq.~(\ref{eq12}), incorporating the estimated microlens shift and focal length. To match the measured profiles, the simulated curves in Fig.~\ref{fig4}e and Fig.~\ref{fig5} were scaled by a calibration factor $C_{\mathrm{cali}}$ of~0.5 applied to the raw simulation output. Accordingly, the curves in Fig.~\ref{fig4}e represent the mapping $C_{\mathrm{cali}}\mathcal{G}{\mathrm{sim}}(\cdot)$ in Eq.~(\ref{cali_eff}). The calibration factor was determined empirically by comparing the simulated phase-gradient amplitudes with the experimental measurements and adjusting the scale until the best agreement was reached. This factor indicates that the sensor exhibits lower phase-gradient sensitivity than predicted. One potential contributing factor is a shorter distance between the photosensitive layer and the microlens than anticipated, although achieving such a reduction is challenging in practice. A more likely explanation is that the actual focal spot shape deviates from our simplified model, such that for the same displacement~$\Delta x$, the resulting normalized differential signal is smaller.

As the actual phase gradient of the microbeads diverges at the bead edges, the simulation computes the gradient by applying finite differentiation with a 2~$\mu\mathrm{m}$ sampling step at the image plane, corresponding to the microlens size of a $2 \times 2$ pixel group. The normalized differential signals exhibit strong agreement with the simulated results after scaling under low-coherence illumination ($\sigma = 1.16$). This confirms the linear relationship between the actual phase gradient and the normalized differential signals at the sensor center, consistent with the simulation. The phase reconstruction is then performed according to Eq.~(\ref{eq13}), following the workflow shown in Fig.~\ref{fig1}c. At $\sigma = 1.16$, the reconstructed phase map matches the theoretical prediction. As $\sigma$ decreases, however, $\bar{\gamma}_x$ and $\bar{\gamma}_y$ become more noisy, and the agreement between the reconstructed phase (derived from these signals) and the ideal spherical gradient shape degrades.

Additionally, we investigated this effect for different immersion media in Fig.~S1, corresponding to various refractive index differences. For microbeads immersed in all tested immersion media, the agreement between the theoretical phase and the reconstructed phase degraded with increasing illumination coherence. Theoretically, higher coherence leads to stronger diffraction, which distorts the reconstructed phase structure of the microbead, and this effect becomes more pronounced  for larger refractive index contrasts. Moreover, high coherence can introduce significant artifacts arising from coherent scattering by out-of-plane structures~\cite{3_tian_quantitative_2015}. These results indicate that using lower-coherence illumination is beneficial for achieving more reliable phase detection, provided that the coherence length remains larger than the size of the $2 \times 2$ pixel group, which ensures repeatability of the phase measurement.

\begin{figure}[htbp]
\centering\includegraphics[width=12.5cm]{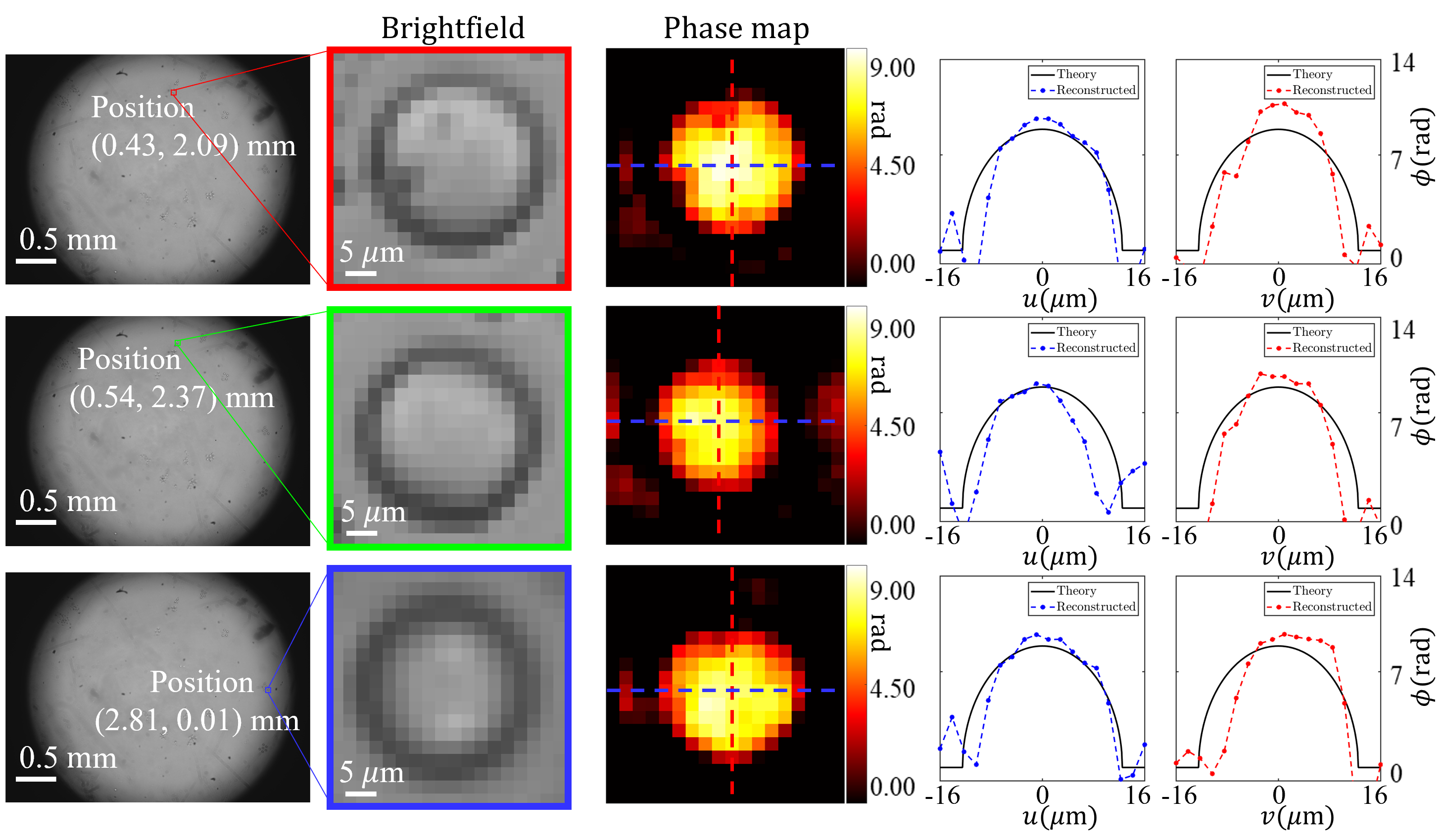}
\caption{Reconstructed phase maps of the microbeads ($n_s = 1.59$) captured at different positions of the imaging FOV. The image is acquired under a coherence parameter $\sigma = 1.16$ with an environmental refractive index $n_0 = 1.56$. The microbead positions are indicated on the left, with the central pixel defined as the origin. The bright-field images are generated by binning all pixels within each $2 \times 2$ pixel group under the same microlens and merging all color channels.}
\label{fig6}
\end{figure}

Figure~\ref{fig6} shows the reconstructed phase gradients at different positions on the sensor. As indicated in Fig.~\ref{fig4}e, the same phase gradient is expected to produce smaller $\bar{\gamma}_x$ and $\bar{\gamma}_y$ values at positions farther from the sensor center along their corresponding directions. Detailed distributions of the normalized differential signals of the microbeads in Fig.~ \ref{fig6} are provided in Supplementary Fig.~S2, which confirms this trend. The phase gradients employed for phase reconstruction are therefore estimated based on the relationship illustrated in Fig.~\ref{fig4}e at the respective positions. The experimental results demonstrate that the sensor maintains relatively high stability in phase-gradient measurements at positions away from the central FOV, and the reconstructed phase maps still retain a spherical-like profile.

\begin{figure}[htbp]
\centering\includegraphics[width=12.5cm]{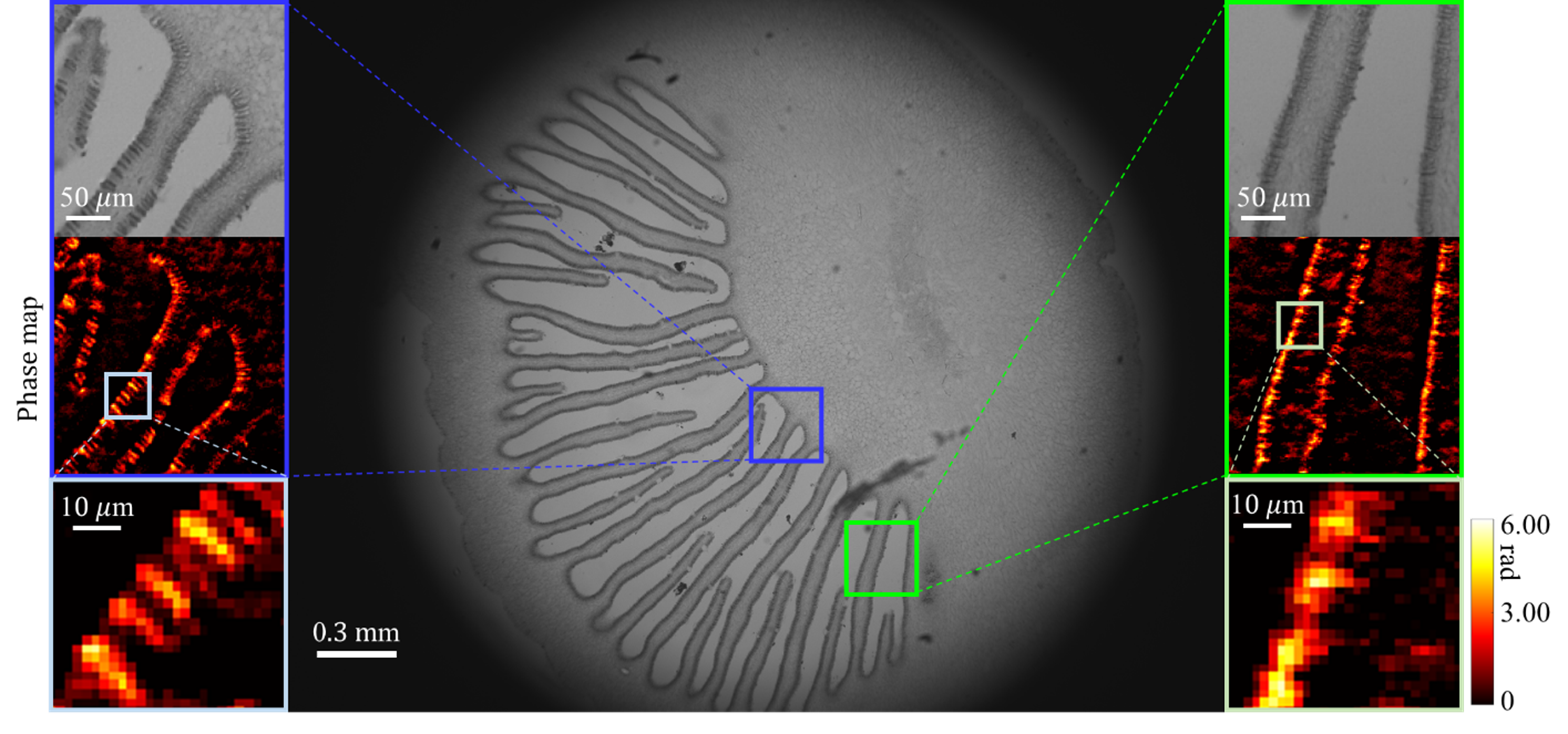}
\caption{Microscopic image and reconstructed phase map of a mushroom slice (\textit{Coprinus} c.s.). The image is acquired with a coherence parameter $\sigma = 1.16$. The central image presents the bright-field image, generated by binning all pixels within each $2 \times 2$ pixel group under the same microlens and combining all color channels. The blue box highlights the region near the center of the imaging FOV, while the green box indicates the region deviating from the center.}
\label{fig7}
\end{figure}

Figure~\ref{fig7} shows an image of a mushroom slide (\textit{Coprinus} c.s.) and the corresponding phase reconstruction obtained by QP$^{2}$GI with the setup configured as in Fig.~\ref{fig3}a. Each gill (\textit{lamella}) of the sample contains small, ellipsoidal basidiospores ($\sim$10~$\mu\mathrm{m}$ in size), which are suitable for analyzing the phase gradients in both the horizontal and vertical directions.  

For each region, the normalized differential signal is derived by subtracting the background from the measured differential signal. The background is estimated by applying an average filter to the original differential signals in Eq.~(\ref{eq6}), with a kernel size of 17. The reconstructed phase maps in both the central region (blue box) and the off-center region (green box) reveal the structural features of the basidiospores.  However, as illustrated in Fig. \ref{fig3}c, the background becomes larger and more nonlinear at positions farther from the center—rendering it increasingly difficult to remove. Such background not only induces degradation of the reconstructed phase map but also narrows the reliable phase detection range, since the focal-spot shift under each microlens should remain within the valid region. This highlights the importance of achieving better CRA matching between the optical system and the sensor, which can reduce position-dependent background variations, mitigating artifacts in normalized differential signals, and ultimately enhancing the accuracy of phase reconstruction across the entire FOV.

\section{Conclusions}

In conclusion, we have demonstrated QP$^{2}$GI, a single-shot quantitative phase imaging technique using a commercially available quad-pixel PDAF sensor and simple partially coherent LED illumination. We have established a propagation model based on $2 \times 2$ pixel groups covered by a single microlens and validated its applicability to microscopic systems. Theoretically, the maximum measurable phase gradient is jointly constrained by the NA of the imaging system and by the microlens aperture corresponding to its $2 \times 2$ pixel group. Although each sub-aperture used for phase detection is smaller than that of a traditional Shack--Hartmann wavefront sensor (typically 100–300~$\mu\mathrm{m}$~\cite{25_neal_shack-hartmann_2002,32_tyson_principles_2022}), the quad-pixel PDAF sensor benefits from a much shorter distance between the microlens and the photosensitive layer, which allows for a potentially broader measurable range of phase gradient. Importantly, the single-shot acquisition capability of QP$^{2}$GI enables quantitative phase imaging at the full frame rate of the sensor, offering a path toward high-speed dynamic phase imaging.This is particularly advantageous for capturing rapid biological processes (e.g., cell dynamics and intracellular transport), which are often challenging to observe with multi-frame or scanning-based QPI methods. By leveraging the native architecture of quad-pixel PDAF sensors, QP$^{2}$GI could further extend the dynamic imaging capability of QPI~\cite{GT_CasteleiroCosta2022DynamicQOBM,DQPI_Krizova2015DynamicPhaseQPI}, especially in applications requiring high temporal resolution. Furthermore, capturing a rapid sequence of frames at different focal planes, the quad-pixel PDAF sensor can also enable high-speed three-dimensional quantitative phase imaging. Phase-gradient maps acquired at different focal depths can be combined via computational  three-dimensional reconstruction algorithms~\cite{3d_qpi1_Jenkins:15,3d_qpi2_Chen:16} to recover volumetric refractive-index distributions, thereby extending QP$^{2}$GI from two-dimensional phase mapping to rapid volumetric phase imaging of dynamic biological processes.

The sensor exhibits superior performance relatively under low coherence ($\sigma = 1.16$), which suppresses artifacts associated with high coherence. The phase map of a 25~$\mu\mathrm{m}$ microbead with a refractive index contrast of $\Delta n = 0.03$ at $\sigma = 1.16$ can be reconstructed with reasonable precision via single-shot acquisition. Both theoretical and experimental analyses confirm that phase detection remains repeatable, provided that the coherence length is sufficiently larger than the size of a $2 \times 2$ pixel group.

The sensor’s phase-gradient detection capability also degrades in regions farther from the sensor center, attributed to CRA mismatch. Notably, the experimental results still demonstrate relatively stable phase detection in peripheral regions. This robustness is likely due to the focal spot projected onto each $2 \times 2$ pixel group being larger in practice than predicted by the simulation. A larger focal spot reduces the variance in the normalized intensity difference across pixels as the spot position shifts, thereby  decreasing the sensor’s sensitivity to phase gradients but extending the measurable range and enhancing the robustness of phase detection in peripheral regions. This also explains why the simulated differential signals must be scaled by a factor of 0.5 to match the experimental results. A more accurate simulation would necessitate detailed information regarding the microlens geometry and arrangement.

For biological samples with finer structures—e.g., \textit{Coprinus} (c.s.), which contains small, ellipsoidal basidiospores—the reconstructed phase maps still retain the overall structural features of the sample’s phase. However, the reconstruction quality degrades in peripheral regions, where CRA mismatch induces larger and more nonlinear background variations. This highlights the critical importance of CRA matching between the sensor and the imaging system for ensuring reliable phase measurements.

Multi-camera array microscopes (MCAM)~\cite{33_harfouche_imaging_2023,34_kreiss_recording_2025}, which employ high-NA finite-conjugate lenses with relatively low magnification (0.1–0.3) to maintain overlap between each sub-FOV, inherently require relatively large CRA. This renders them strong candidates for integration with such sensors. Furthermore, CRA matching between the optical system and the sensor can be further improved via the use of a relay lens, which modifies the chief-ray distribution at the sensor plane and provides a practical pathway to enhanced phase measurement accuracy in microscopic systems employing this type of sensor.

Another potential approach is to design a quad-pixel PDAF sensor with minimal intentional microlens shift, thereby achieving intrinsic CRA matching to microscopic systems with small CRA via the sensor architecture. In this configuration, the optical axis of each microlens aligns with that of the imaging system, effectively eliminating CRA-induced phase detection bias, simplifying system calibration, and providing a more robust and inherently compatible platform for QP$^{2}$GI.

\begin{backmatter}
\bmsection{Funding}

This work was not supported by any funding agency.

\bmsection{Disclosures}

X.B. was a student at Duke University during this work and is currently employed by Ramona Optics, Inc.  
R.H. is both a faculty member at Duke University and a co-founder of Ramona Optics, Inc.    
The remaining authors declare no conflicts of interest.

\bmsection{Data availability}

The data that support the findings of this study are available upon reasonable request from the authors.
\bmsection{Supplemental document}

See Supplement~1 for supporting content.

\end{backmatter}

\bibliography{quad_dpc}

\begin{thebibliography}{10}
\newcommand{\enquote}[1]{``#1''}

\bibitem{QPI1_nguyen_quantitative_2022}
T.~L. Nguyen, S.~Pradeep, R.~L. Judson-Torres, \emph{et~al.}, \enquote{Quantitative {Phase} {Imaging}: {Recent} {Advances} and {Expanding} {Potential} in {Biomedicine},} {\protect\JournalTitle{ACS Nano}} \textbf{16}, 11516--11544 (2022).

\bibitem{QPI2_Majeed}
H.~Majeed, S.~Sridharan, M.~Mir, \emph{et~al.}, \enquote{Quantitative phase imaging for medical diagnosis,} {\protect\JournalTitle{J. Biophotonics}} \textbf{10}, 177--205 (2017).

\bibitem{HOLO_2_huang_quantitative_2024}
Z.~Huang and L.~Cao, \enquote{Quantitative phase imaging based on holography: trends and new perspectives,} {\protect\JournalTitle{Light Sci Appl}} \textbf{13}, 145 (2024). Publisher: Nature Publishing Group.

\bibitem{HOLO1_mann_high-resolution_2005}
C.~J. Mann, L.~Yu, C.-M. Lo, and M.~K. Kim, \enquote{High-resolution quantitative phase-contrast microscopy by digital holography,} {\protect\JournalTitle{Opt. Express}} \textbf{13}, 8693--8698 (2005).

\bibitem{6_sheppard_defocused_2004}
C.~J. Sheppard, \enquote{Defocused transfer function for a partially coherent microscope and application to phase retrieval,} {\protect\JournalTitle{Journal of the Optical Society of America A}} \textbf{21}, 828--831 (2004).

\bibitem{7_bao_two_2019}
Y.~Bao and T.~K. Gaylord, \enquote{Two improved defocus quantitative phase imaging methods: discussion,} {\protect\JournalTitle{Journal of the Optical Society of America A}} \textbf{36}, 2104--2114 (2019).

\bibitem{1_lin_quantitative_2018}
Y.-Z. Lin, K.-Y. Huang, and Y.~Luo, \enquote{Quantitative differential phase contrast imaging at high resolution with radially asymmetric illumination,} {\protect\JournalTitle{Optics Letters}} \textbf{43}, 2973--2976 (2018).

\bibitem{2_fan_optimal_2019}
Y.~Fan, J.~Sun, Q.~Chen, \emph{et~al.}, \enquote{Optimal illumination scheme for isotropic quantitative differential phase contrast microscopy,} {\protect\JournalTitle{Photon. Res.}} \textbf{7}, 890--904 (2019).

\bibitem{3_tian_quantitative_2015}
L.~Tian and L.~Waller, \enquote{Quantitative differential phase contrast imaging in an {LED} array microscope,} {\protect\JournalTitle{Opt. Express}} \textbf{23}, 11394--11403 (2015).

\bibitem{5_hamilton_improved_1984}
D.~K. Hamilton, C.~J.~R. Sheppard, and T.~Wilson, \enquote{Improved imaging of phase gradients in scanning optical microscopy,} {\protect\JournalTitle{Journal of Microscopy}} \textbf{135}, 275--286 (1984).

\bibitem{FPM1_zheng_wide-field_2013}
G.~Zheng, R.~Horstmeyer, and C.~Yang, \enquote{Wide-field, high-resolution fourier ptychographic microscopy,} {\protect\JournalTitle{Nat. Photonics}} \textbf{7}, 739--745 (2013).

\bibitem{FPM2_ou_quantitative_2013}
X.~Ou, R.~Horstmeyer, C.~Yang, and G.~Zheng, \enquote{Quantitative phase imaging via fourier ptychographic microscopy,} {\protect\JournalTitle{Opt. Lett.}} \textbf{38}, 4845--4848 (2013).

\bibitem{9_levoy_light_2006}
M.~Levoy, R.~Ng, A.~Adams, \emph{et~al.}, \enquote{Light field microscopy,} in \emph{ACM SIGGRAPH 2006 Papers,}  (ACM Press, 2006), p. 924.

\bibitem{10_broxton_wave_2013}
M.~Broxton, L.~Grosenick, S.~Yang, \emph{et~al.}, \enquote{Wave optics theory and 3-{D} deconvolution for the light field microscope,} {\protect\JournalTitle{Opt. Express}} \textbf{21}, 25418--25439 (2013).

\bibitem{11_bimber_light-field_2019}
O.~Bimber and D.~Schedl, \enquote{Light-field microscopy: A review,} {\protect\JournalTitle{J. Neurol. Neuromed.}} \textbf{4}, 1--6 (2019).

\bibitem{12_mignard-debise_light-field_2015}
L.~Mignard-Debise and I.~Ihrke, \enquote{Light-field microscopy with a consumer light-field camera,} in \emph{2015 {International} {Conference} on {3D} {Vision},}  (2015), pp. 335--343.

\bibitem{13_platt_history_2001}
B.~C. Platt and R.~Shack, \enquote{History and {Principles} of {Shack}-{Hartmann} {Wavefront} {Sensing},} {\protect\JournalTitle{Journal of Refractive Surgery}} \textbf{17}, S573--S577 (2001).

\bibitem{14_lane_wave-front_1992}
R.~G. Lane and M.~Tallon, \enquote{Wave-front reconstruction using a {Shack}–{Hartmann} sensor,} {\protect\JournalTitle{Appl. Opt.}} \textbf{31}, 6902--6908 (1992).

\bibitem{15_primot_theoretical_2003}
J.~Primot, \enquote{Theoretical description of {Shack}–{Hartmann} wave-front sensor,} {\protect\JournalTitle{Optics Communications}} \textbf{222}, 81--92 (2003).

\bibitem{16_prieto_analysis_2000}
P.~M. Prieto, F.~Vargas-Mart\'{i}n, S.~Goelz, and P.~Artal, \enquote{Analysis of the performance of the {Hartmann}–{Shack} sensor in the human eye,} {\protect\JournalTitle{J. Opt. Soc. Am. A}} \textbf{17}, 1388--1398 (2000).

\bibitem{17_wei_design_2010}
X.~Wei and L.~Thibos, \enquote{Design and validation of a scanning {Shack} {Hartmann} aberrometer for measurements of the eye over a wide field of view,} {\protect\JournalTitle{Opt. Express}} \textbf{18}, 1134--1143 (2010).

\bibitem{18_jeong_measurement_2005}
T.~M. Jeong, M.~Menon, and G.~Yoon, \enquote{Measurement of wave-front aberration in soft contact lenses by use of a {Shack}–{Hartmann} wave-front sensor,} {\protect\JournalTitle{Appl. Opt.}} \textbf{44}, 4523--4527 (2005).

\bibitem{19_abdelazeem_characterization_2023}
R.~M. Abdelazeem, M.~M.~A. Ahmed, and M.~Agour, \enquote{Characterization of thick and contact lenses using an adaptive {Shack}–{Hartmann} wavefront sensor: {Limitations} and solutions,} {\protect\JournalTitle{Optik}} \textbf{283}, 170922 (2023).

\bibitem{20_dayton_atmospheric_1992}
D.~Dayton, B.~Pierson, B.~Spielbusch, and J.~Gonglewski, \enquote{Atmospheric structure function measurements with a {Shack}–{Hartmann} wave-front sensor,} {\protect\JournalTitle{Opt. Lett.}} \textbf{17}, 1737--1739 (1992).

\bibitem{21_kikuchi_125_2024}
K.~Kikuchi, K.~Tomioka, T.~Usui, \emph{et~al.}, \enquote{A 1.25~$\mu$m, 59.3~mpixel, 60~fps cmos image sensor with 2$\times$1 multi-directional phase-detection pixels,} in \emph{2024 {IEEE} {SENSORS},}  (2024), pp. 1--4.

\bibitem{22_fukuda_compressed_nodate}
K.~Fukuda, \enquote{A compressed $n\times n$ multi-pixel imaging and cross phase-detection af with $n\times1$ rgrb + $1\times$~ngb hetero multi-pixel image sensors,} in \emph{Proc. Int. Image Sensor Workshop,}  (2021), p. R28.

\bibitem{23_noauthor_ov50a_nodate}
\enquote{{OV50A},} https://www.ovt.com/products/ov50a/.

\bibitem{24_jutamulia_phase_2022}
S.~Jutamulia, \enquote{Phase detection autofocus ({PDAF}) image sensor as applied to {3D} imaging,} {\protect\JournalTitle{Journal of Physics: Conference Series}} \textbf{2274}, 012005 (2022).

\bibitem{25_neal_shack-hartmann_2002}
D.~R. Neal, J.~Copland, and D.~A. Neal, \enquote{Shack-{Hartmann} wavefront sensor precision and accuracy,} in \emph{Advanced Characterization Techniques for Optical, Semiconductor, and Data Storage Components,}  vol. 4779 (Seattle, WA, 2002), pp. 148--160.

\bibitem{26_thomas_comparison_2006}
S.~Thomas, T.~Fusco, A.~Tokovinin, \emph{et~al.}, \enquote{Comparison of centroid computation algorithms in a {Shack}–{Hartmann} sensor,} {\protect\JournalTitle{Monthly Notices of the Royal Astronomical Society}} \textbf{371}, 323--336 (2006).

\bibitem{27_goodman_statistical_2015}
J.~W. Goodman, \emph{Statistical {Optics}} (John Wiley \& Sons, 2015).

\bibitem{28_taflove_computational_2005}
A.~Taflove, S.~C. Hagness, and M.~Piket-May, \enquote{Computational electromagnetics: the finite-difference time-domain method,} {\protect\JournalTitle{The Electrical Engineering Handbook}} \textbf{3}, 15 (2005).

\bibitem{29_noauthor_method_nodate}
R.~Frankot and R.~Chellappa, \enquote{A method for enforcing integrability in shape from shading algorithms,} {\protect\JournalTitle{IEEE Transactions on Pattern Analysis and Machine Intelligence}} \textbf{10}, 439--451 (1988).

\bibitem{30_holland_development_1997}
S.~Holland, N.~Wang, and W.~Moses, \enquote{Development of low noise, back-side illuminated silicon photodiode arrays,} {\protect\JournalTitle{IEEE Transactions on Nuclear Science}} \textbf{44}, 443--447 (1997).

\bibitem{31_taverni_front_2018}
G.~Taverni, D.~Paul~Moeys, C.~Li, \emph{et~al.}, \enquote{Front and back illuminated dynamic and active pixel vision sensors comparison,} {\protect\JournalTitle{IEEE Transactions on Circuits and Systems II: Express Briefs}} \textbf{65}, 677--681 (2018).

\bibitem{32_tyson_principles_2022}
R.~K. Tyson and B.~W. Frazier, \emph{Principles of {Adaptive} {Optics}} (CRC Press, 2022), 5th ed.

\bibitem{GT_CasteleiroCosta2022DynamicQOBM}
C.~P. Casteleiro, B.~Wang, C.~E. Serafini, \emph{et~al.}, \enquote{Functional imaging with dynamic quantitative oblique back-illumination microscopy,} {\protect\JournalTitle{J. Biomed. Opt.}} \textbf{27}, 066502 (2022).

\bibitem{DQPI_Krizova2015DynamicPhaseQPI}
A.~Krizova, J.~Collakova, Z.~Dostal, \emph{et~al.}, \enquote{Dynamic phase differences based on quantitative phase imaging for the objective evaluation of cell behavior,} {\protect\JournalTitle{J. Biomed. Opt.}} \textbf{20}, 111214 (2015).

\bibitem{3d_qpi1_Jenkins:15}
M.~H. Jenkins and T.~K. Gaylord, \enquote{Three-dimensional quantitative phase imaging via tomographic deconvolution phase microscopy,} {\protect\JournalTitle{Appl. Opt.}} \textbf{54}, 9213--9227 (2015).

\bibitem{3d_qpi2_Chen:16}
M.~Chen, L.~Tian, and L.~Waller, \enquote{3d differential phase contrast microscopy,} {\protect\JournalTitle{Biomed. Opt. Express}} \textbf{7}, 3940--3950 (2016).

\bibitem{33_harfouche_imaging_2023}
M.~Harfouche, K.~Kim, K.~C. Zhou, \emph{et~al.}, \enquote{Imaging across multiple spatial scales with the multi-camera array microscope,} {\protect\JournalTitle{Optica}} \textbf{10}, 471--480 (2023).

\bibitem{34_kreiss_recording_2025}
L.~Kreiss, W.~Tang, R.~Balla, \emph{et~al.}, \enquote{Recording dynamic facial micro-expressions with a multi-focus camera array,} {\protect\JournalTitle{Biomed. Opt. Express}} \textbf{16}, 617--627 (2025).

\end{thebibliography}

\newpage

\vspace*{1em}
\begin{center}
    {\LARGE\textbf{Towards a mobile quantitative phase imaging microscope with smartphone phase-detection sensors}}\\[0.5em]
    \rule{0.5\textwidth}{0.4pt}\\[0.5em]
    {\large Supplement 1}\\[1em] 
\end{center}

\renewcommand\thesection{S\arabic{section}}
\setcounter{section}{0}
\section{Additional Images}

\renewcommand{\thefigure}{S\arabic{figure}}
\setcounter{figure}{0} 
\begin{figure}[htbp]
\centering\includegraphics[width=12.5cm]{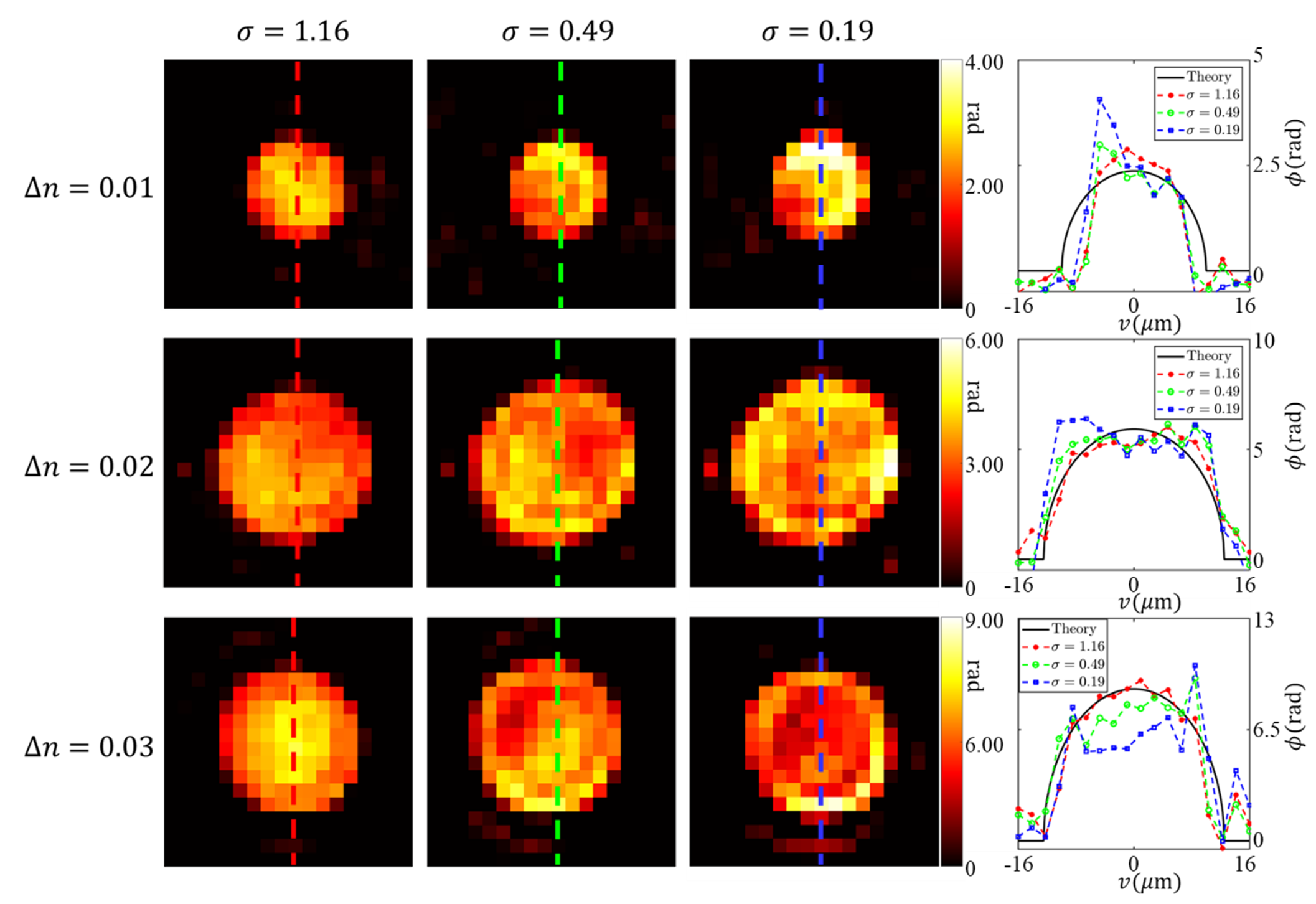}
\caption{Reconstructed phase maps of the microbeads (refractive index $n_s = 1.59$) captured at the central field of view (FOV). 
The microbeads (nominal diameter = $25~\mu\text{m}$) are immersed in environments with refractive indices $n_0 = 1.58$, $1.57$, and $1.56$, respectively. 
The results shown in each row are reconstructed from the same microbead, acquired under identical conditions except for the illumination coherence. 
The microbead in the $n_0 = 1.58$ medium exhibits a slightly smaller actual diameter of $\sim 22~\mu\text{m}$. 
Illumination with coherence parameters $\sigma = 1.16$, $0.49$, and $0.19$ is applied during imaging. The coherence parameter is defined as the ratio of the illumination NA to the objective NA.
The rightmost column presents the central profiles of the reconstructed phase, compared with the corresponding theoretical predictions.}
\label{figs1}
\end{figure}

\begin{figure}[htbp]
\centering\includegraphics[width=12.5cm]{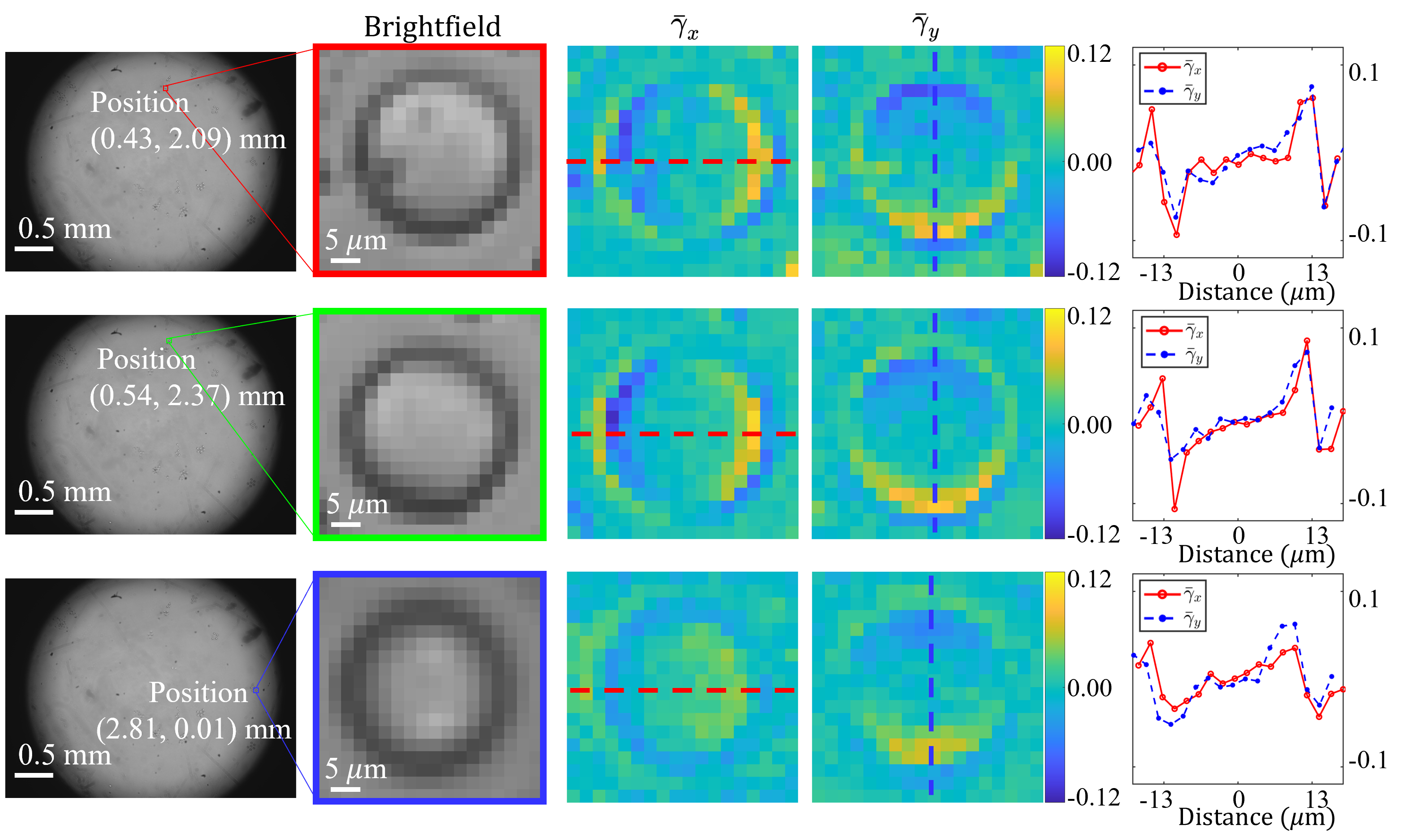}
\caption{Normalized differential signals $\bar{\gamma}_x$ and $\bar{\gamma}_y$ of the microbeads ($n_s = 1.59$) captured at different positions of the imaging FOV. 
The images are acquired under a coherence parameter $\sigma = 1.16$ with an environmental refractive index of $n_0 = 1.56$. 
The microbead positions are indicated in the left column, with the central pixel defined as the origin.
The rightmost column presents the central profiles of the normalized differential signals for microbeads at different sensor positions.
}
\label{figs2}
\end{figure}

\clearpage
\section{Propagation Model}

\begin{figure}[htbp]
\centering
\includegraphics[width=12.5cm]{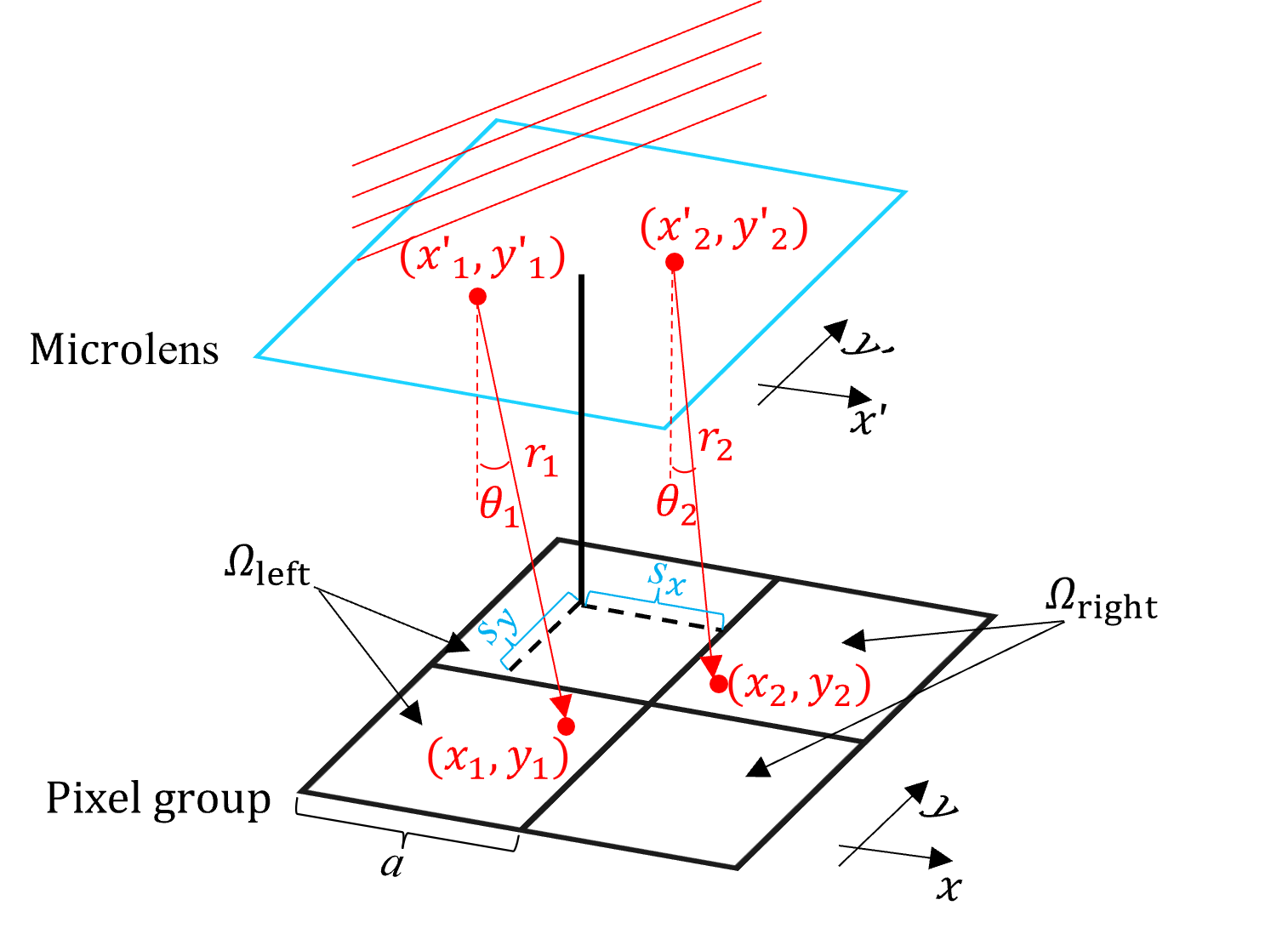}
\caption{Geometry of mutual-coherence propagation within a single $2\times2$ pixel group.}
\label{figs3}
\end{figure}

The global coordinates on the sample plane are denoted as $(u, v)$, and those on the image plane as $(u', v')$. 
For each $2\times2$ pixel group, the local coordinates within a pixel are represented by $(x, y)$, 
and the corresponding coordinates on the microlens as $(x', y')$. To characterize the field distribution beneath each $2\times2$ pixel group, we model the 
propagation from the microlens plane to the pixel plane using the Huygens--Fresnel integral. 
The complete expression used in our simulation, evaluated at the coincident point $(x_1,y_1)=(x_2,y_2)=(x,y)$, is given by:

\setcounter{equation}{0}
\renewcommand\theequation{S\arabic{equation}}

\begin{equation}
\begin{aligned}
I(x,y;u',v') &=
\int_{-a+s_x(u',v')}^{a+s_x(u',v')}
\!\!\int_{-a+s_y(u',v')}^{a+s_y(u',v')}
\!\!\int_{-a+s_x(u',v')}^{a+s_x(u',v')}
\!\!\int_{-a+s_y(u',v')}^{a+s_y(u',v')}
J(x_1',y_1';x_2',y_2') \\
&\quad \times 
M^*(x_1',y_1')\,M(x_2',y_2') \\
&\quad \times 
\psi^*(x_1',y_1';u',v')\,\psi(x_2',y_2';u',v') \\
&\quad \times
\Theta^*(x_1',y_1';u',v')\,\Theta(x_2',y_2';u',v') \\
&\quad \times 
\exp\!\left[-j\frac{2\pi}{\bar{\lambda}}(r_2-r_1)\right]
\frac{\cos(\theta_1)\,\cos(\theta_2)}
{\bar{\lambda}^2 r_1 r_2}\,
dx_1' dy_1' dx_2' dy_2'.
\end{aligned}
\label{eq10_sup}
\end{equation}

The phase terms in Eq.~(\ref{eq10_sup}) account for the contributions from the sample's 
phase gradient and the local chief-ray angle (CRA), and are given by:

\begin{equation}
\begin{aligned}
\psi(x',y';u',v') &=
\exp\!\left[-j\left(
\frac{\partial\phi(u',v')}{\partial u'} x' + \frac{\partial\phi(u',v')}{\partial v'}y'
\right)\right], \\[4pt]
\Theta(x',y';u',v') &=
\exp\!\left[-jk\left(
\theta_{r,x}(u',v')x' + \theta_{r,y}(u',v')y'
\right)\right].
\end{aligned}
\label{eqPsiTheta_sup}
\end{equation}

Here, $a$ denotes the pixel width, and the microlens aperture is assumed to match the pixel size. 
The quantities $s_x(u',v')$ and $s_y(u',v')$ represent the lateral microlens shifts across the FOV, 
while $\theta_{r,x}(u',v')$ and $\theta_{r,y}(u',v')$ denote the corresponding CRA components in different directions. The geometric distances $r_1$, $r_2$ and angles $\theta_1$, $\theta_2$ describe 
the propagation from the microlens plane to the pixel plane. $J(x'_1,y'_1; x'_2,y'_2)$ denotes the mutual intensity function evaluated on the microlens plane. The microlens is modeled as an ideal thin 
lens via the complex transmittance $M(x',y')=\exp[-\,j\,\frac{\pi}{\bar{\lambda}f_l}\,(x'^2 + y'^2)]$, where $f_l$ denotes the microlens focal length.

To compute the directional signals used in Eq.~(\ref{eq6}), we integrate the simulated 
intensity over the corresponding regions of a $2\times2$ pixel group:

\begin{equation}
\begin{aligned}
I^{\mathrm{sim}}_{\mathrm{left}}(u',v')  &= \iint_{\Omega_{\mathrm{left}}} I(x,y;u',v')\,dx\,dy, \\
I^{\mathrm{sim}}_{\mathrm{right}}(u',v') &= \iint_{\Omega_{\mathrm{right}}} I(x,y;u',v')\,dx\,dy, \\
I^{\mathrm{sim}}_{\mathrm{up}}(u',v')    &= \iint_{\Omega_{\mathrm{up}}}    I(x,y;u',v')\,dx\,dy, \\
I^{\mathrm{sim}}_{\mathrm{low}}(u',v')  &= \iint_{\Omega_{\mathrm{low}}}  I(x,y;u',v')\,dx\,dy.
\end{aligned}
\label{simu_intensity_sup}
\end{equation}

For typical LED illumination, the coherence length exceeds the $\sim 2~\mu$m width of a pixel 
group. Under this condition, the mutual intensity approaches the fully coherent limit within a pixel group, and 
Eq.~(\ref{eq10_sup}) simplifies to the coherent form:

\begin{equation}
\begin{aligned}
I(x,y;u',v') &=
\left|
\int_{-a+s_x(u',v')}^{a+s_x(u',v')}
\int_{-a+s_y(u',v')}^{a+s_y(u',v')}
M(x',y') \, \psi(x',y';u',v') \, \Theta(x',y';u',v')
\right. \\
&\quad\left.
\times \exp\!\left[-j\frac{2\pi}{\bar{\lambda}}r\right]
\frac{\cos(\theta)}{\bar{\lambda} r} 
\, dx' \, dy'
\right|^2 .
\end{aligned}
\label{eq11_sup}
\end{equation}

\end{document}